\documentclass[12pt,twoside,english,nofootinbib,preprint,showpacs,preprintnumbers,floatfix]{revtex4}
\usepackage{mathptmx}
\usepackage{helvet}
\usepackage{courier}
\usepackage[T1]{fontenc}
\usepackage[latin9]{inputenc}
\usepackage[a4paper]{geometry}
\usepackage{amsmath}
\usepackage{amssymb}
\usepackage{graphicx}
\usepackage{esint}

\makeatletter
\@ifundefined{textcolor}{}
{%
 \definecolor{BLACK}{gray}{0}
 \definecolor{WHITE}{gray}{1}
 \definecolor{RED}{rgb}{1,0,0}
 \definecolor{GREEN}{rgb}{0,1,0}
 \definecolor{BLUE}{rgb}{0,0,1}
 \definecolor{CYAN}{cmyk}{1,0,0,0}
 \definecolor{MAGENTA}{cmyk}{0,1,0,0}
 \definecolor{YELLOW}{cmyk}{0,0,1,0}
}

\usepackage{babel}

\usepackage{babel}

\usepackage{babel}

\makeatother

\usepackage{babel}
\begin{document}
\begin{flushright}
TTK-14-14 
\par\end{flushright}

\vspace{4mm}

\begin{center}
\textbf{\LARGE Determination of the Higgs CP mixing angle}{\LARGE {}
} 
\par\end{center}

\begin{center}
\textbf{\LARGE in the tau decay channels at the LHC}{\LARGE{} }
\par\end{center}{\LARGE \par}

\begin{center}
\textbf{\LARGE including the Drell-Yan background}{\LARGE{} }
\par\end{center}{\LARGE \par}

\begin{center}
\vspace{6mm}

\par\end{center}

\begin{center}
\textbf{\large Stefan Berge}{\large }%
\footnote{\texttt{\small berge@physik.rwth-aachen.de}%
}\textbf{\large ,}{\large{} }\textbf{\large Werner Bernreuther}{\large }%
\footnote{\texttt{\small breuther@physik.rwth-aachen.de}%
}\textbf{\large {} and Sebastian Kirchner}{\large }%
\footnote{\texttt{\small kirchner@physik.rwth-aachen.de}%
}\textbf{\large {} }{\large{} }
\par\end{center}{\large \par}

\begin{center}
Institut f\"ur Theoretische Physik, RWTH Aachen University, 52056 Aachen,
Germany 
\par\end{center}

\begin{center}
\vspace{10mm}
 \textbf{Abstract} 
\parbox[t]{\textwidth}
{\small{ 
  We investigate how precisely the CP nature of the 125 GeV
Higgs boson resonance $h$ can be
 unraveled at the LHC in its decays to $\tau$ pairs, $h \to \tau^-\tau^+$.
  We use a method
  which allows to  determine the scalar-pseudoscalar Higgs  mixing
  angle $\phi_\tau$ in this decay mode.
  This mixing angle can be extracted from the distribution of a signed angle, denoted by $\varphi_{CP}^{*}$,
  which we analyze for the major charged-prong $\tau$ decays. For definiteness,
 we consider Higgs-boson production by gluon fusion at NLO QCD. We
 take into account  also
  the irreducible  background from Drell-Yan production,
  $Z^*/\gamma^*\to\tau\tau$, at NLO QCD. We compute, for the signal and background
   reactions, angular and energy
   correlations of the charged prongs  and analyze which type of cuts  suppress the
   Drell-Yan background. An important feature of this background is
   that its contribution to the  distribution of our observable $\varphi_{CP}^{*}$ is a
   flat line, also at NLO QCD. By separating the Drell-Yan $\tau$
   events into two different sets, two different non-trivial 
  $\varphi_{CP}^{*}$ distributions are obtained. Based on this observation
      we propose to use these sets for calibation purposes.
    By Monte Carlo simulation we study also the
   effect of measurement uncertainties on this distribution.
  We  estimate that the Higgs
 mixing angle $\phi_\tau$ can be determined with our method to a
 precision of $\Delta\phi_\tau\simeq 14^\circ$ $(5^\circ)$ at the high
 luminosity LHC (14 TeV)
 with an integrated luminosity of $500$ fb$^{-1}$ (3 ab$^{-1}$).
 }}

\par\end{center}

\vspace{10mm}
PACS numbers: 11.30Er, 12.60.Fr, 14.80.Bn \\
 Keywords: Higgs bosons, tau leptons, parity, Z boson, spin correlations,
CP violation\newpage{}

\section{Introduction}

The Large Hadron Collider (LHC) has had its first major triumph with
the discovery of a new, electrically neutral boson $h$ with mass
$m_{h}\simeq125$ GeV by the ATLAS and CMS experiments \cite{Aad:2012tfa,Chatrchyan:2012ufa}.
According to present experimental knowledge on the spin and parity
of $h$ \cite{Chatrchyan:2012jja,Aad:2013xqa} and its couplings to
gauge bosons and quarks and leptons \cite{Chatrchyan:2012jja,Aad:2013xqa,Chatrchyan:2013lba,Aad:2013wqa,Chatrchyan:2013mxa,ATLtauconf,Chatrchyan:2014vua,Chatrchyan:2014nva,ATLHprconf},
the properties of this boson agree with those of the Higgs boson predicted
by the Standard Model (SM) of particle physics. In particular, the
decay of $h$ to a pair of $\tau$ leptons was recently established
\cite{ATLtauconf,Chatrchyan:2014nva}.

Nevertheless, much more experimental analysis is required for completely
unraveling the properties of this resonance. In particular, although
the LHC data strongly prefer that $h$ is a $J^{P}=0^{+}$ state,
it is not yet excluded with high probability that $h$ has a pseudoscalar
component. How the spin and $CP$ quantum number of a neutral Higgs-like
boson can be pinned down at a hadron collider or at a future (linear)
$e^{+}e^{-}$ collider  has been investigated in numerous papers,
including  \cite{Dell'Aquila:1985vc,Dell'Aquila:1988rx,Bernreuther:1993df,Bernreuther:1993hq,Soni:1993jc,Chang:1993jy,Barger:1993wt,Kramer:1993jn,Hagiwara:1993sw,Arens:1994nc,Skjold:1994qn,Arens:1994wd,Skjold:1995jp,BarShalom:1995jb,Grzadkowski:1995rx,Gunion:1996vv,Bernreuther:1997af,BarShalom:1997sx,Grzadkowski:1999ye,Hagiwara:2000tk,Han:2000mi,Plehn:2001nj,Choi:2002jk,Bower:2002zx,Desch:2003mw,Asakawa:2003dh,Desch:2003rw,Buszello:2002uu,Godbole:2004xe,Rouge:2005iy,Biswal:2005fh,Ellis:2005ika,Accomando:2006ga,Godbole:2007cn,Bhupal Dev:2007is,Klamke:2007cu,Berge:2008wi,Dutta:2008bh,Biswal:2008tg,Berge:2008dr,Reinhard:2009,Christensen:2010pf,DeRujula:2010ys,Berge:2011ij,Godbole:2011hw,Biswal:2012mp,Ellis:2012xd,Banerjee:2012ez,Hagiwara:2012vz,Djouadi:2013yb,Englert:2013opa,Ananthanarayan:2013cia,Frank:2013gca,Harnik:2013aja,Berge:2013jra,Bishara:2013vya,Ellis:2013yxa,Kaczmarska:2014eoa,Przedzinski:2014pla,Chen:2014ona,Kopp:2014rva,Dolan:2014upa,Demartin:2014fia}.

In this paper we elaborate on a method for determining the $CP$ nature
of a Higgs-like resonance at the LHC in its decays to $\tau$ leptons which has
been developed in a series of papers \cite{Berge:2008wi,Berge:2008dr,Berge:2011ij,Berge:2013jra}
both for Higgs production at the LHC and in $e^{+}e^{-}$ collisions.
Our approach is based on the distribution of a signed angle $\varphi_{CP}^{*}$
between the decay planes of the charged-prong decays $\tau^{-}\to a^{-}$
and $\tau^{+}\to a'^{+}$ in the $a^{-}a'^{+}$ zero-momentum frame.
We apply this method to the 125 GeV resonance $h$. We assume that
$h$ is a mixture of a CP-even and  CP-odd state with the CP-odd
 admixture being smaller than the CP-even one.
This assumption is in accord with
the analysis of present data, cf. for instance \cite{Djouadi:2013qya,Bechtle:2013xfa,Brod:2013cka,Bechtle:2014ewa,Inoue:2014nva}.
 We investigate also the contribution of the irreducible background
 $Z^{*}/\gamma^{*}\to\tau\tau$
 to the  $\varphi_{CP}^{*}$ distribution.
While the signal contribution to this  distribution
shows a characteristic dependence on
$\cos(\varphi_{CP}^{*}-2\phi_\tau)$, where the angle $\phi_\tau$
describes the mixing of the scalar and pseudoscalar 
 Higgs component which couple to $\tau$ leptons (see Sec.~\ref{cross_section}), 
we find that the background contribution is flat. 
 We point out how the background events  $Z^{*}/\gamma^*\to\tau\tau$, which are
 numerous at the LHC, can be used by experiments for calibrating the
 measurements of the distribution of the  angle $\varphi_{CP}^{*}$.
 Moreover, we consider
the two-dimensional helicity-angle and energy distributions of the charged prongs
from $\tau^{\mp}$ decay, which differ for $h\to\tau\tau$ and $Z^{*}/\gamma^{*}\to\tau\tau$
because of the different spins of the bosons, and analyze whether
appropriate cuts can enhance the signal-to-background ratio. Furthermore,
we estimate the precision with which the scalar-pseudoscalar mixing
angle $\phi_\tau$ might be measured at the LHC (14 TeV).

The paper is organized as follows. In the next section we briefly
describe the $\tau$-decay modes and decay density matrices which
are used in this analysis. In Sec.~\ref{sec:dsigmaH_LHC} and ~\ref{Observables}
we exhibit the helicity angle distributions of the charged prongs
from $h\to\tau\tau$ decay at the LHC.
  We recall the definition
of the angle $\varphi_{CP}^{*}$  and its
distribution in $h\to\tau\tau$ with subsequent decays to charged prongs,
 which allows to determine the CP-mixing angle $\phi_\tau$.
In Sec.~\ref{sec:dsigmaZ_LHC} we analyze the helicity and azimuthal angle distributions
of the charged prongs for Drell-Yan production of $\tau^{-}\tau^{+}$
at the LHC. In particular, we elaborate on the  distribution of
 the angle $\varphi_{CP}^{*}$ in  $Z^{*}/\gamma^*\to\tau\tau\to\pi^-\pi^+$.
 The discussion in this section is based on the leading-order
distributions. We have computed these distributions also at next-to-leading
order (NLO) QCD. The results are given in Sec.~\ref{sec:Numerical-Results-LHC}
and are compared with the corresponding distributions for inclusive
Higgs production at the LHC, which we computed at NLO QCD 
by including differential Higgs boson-distributions obtained with  the
computer code MCFM  \cite{Campbell:2010ff}
 into our Monte Carlo program.
   We analyze how appropriate
cuts on the polar angle distributions of the charged prongs, respectively
associated cuts on their energies can reduce the irreducible
background.
 Moreover, we study the impact of measurement uncertainties on these
 distributions by Monte Carlo simulation. 
Finally we estimate the precision with which the scalar-pseudoscalar
mixing angle $\phi_\tau$ might be measured at the LHC (14 TeV). 
We summarize in Sec.~\ref{conclusions}.


\section{Higgs-boson production and decay to $\tau\tau$}
\label{cross_section}

Our method to determine the CP nature of the 125 GeV resonance $h$
in its $\tau\tau$ decays, which will be described in Sec. \ref{Observables},
can be applied to any $h$ production mode, but for definiteness,
we consider $h$ production at the LHC by gluon gluon fusion, 
\begin{equation}
p\, p\to h+\, X\label{lhcggjet}
\end{equation}
We consider the decay mode $h\to\tau^{-}\tau^{+}$ with subsequent
decays 
\begin{equation}
h\to\tau^{-}\tau^{+}\to a^{-}a'^{+}+X\,,\label{phitaudec}
\end{equation}
where $a^{\pm},a'^{\pm}\in\{e^{\pm},\mu^{\pm},\pi^{\pm},a_{1}^{L,T,\pm}\}$
and $X$ denotes neutrinos and $\pi^{0}$. We take into account the
main 1- and 3-charged prong $\tau$ decay modes: 
\begin{eqnarray}
\tau & \to & l+\nu_{l}+\nu_{\tau}\,,\label{taulept}\\
\tau & \to & \pi+\nu_{\tau}\,,\label{taupi}\\
\tau & \to & \rho+\nu_{\tau}\to\pi+\pi^{0}+\nu_{\tau}\,,\label{taurho}\\
\tau & \to & a_{1}+\nu_{\tau}\to\pi+2\pi^{0}+\nu_{\tau}\,,\label{taua1}\\
\tau & \to & a_{1}^{L,T}+\nu_{\tau}\to2\pi^{\pm}+\pi^{\mp}+\nu_{\tau}\,.\label{taua1LT}
\end{eqnarray}
We call the decay mode (\ref{taua1LT}) also `1-prong', because the
4-momentum of $a_{1}^{\pm}$ can be obtained from the measured 4-momenta
of the 3 charged pions. The longitudinal $(L)$ and transverse $(T)$
helicity states of the $a_{1}$ resonance can be separated by using
known kinematic distributions
\cite{Rouge:1990kv,Davi93,Kue95,Stahl:2000aq}.

The dynamics of the above $\tau$ decays is, to the precision relevant
for our purposes, known Standard Model physics.
The interaction of a Higgs boson $h$ of arbitrary $CP$ nature to 
$\tau$ leptons is described by the Yukawa Lagrangian 
\begin{equation}
{\cal L}_{Y} = 
-(\sqrt{2}G_{F})^{1/2} m_{\tau}
\left(a_{\tau}\bar{\tau}\tau+b_{\tau}\bar{\tau}i\gamma_{5}\tau\right) h\,,
\label{YukLa}
\end{equation}
where $G_{F}$ denotes the Fermi constant and $a_{\tau}$, $b_{\tau}$
are the reduced dimensionless $\tau$ Yukawa coupling
constants. Instead of  (\ref{YukLa}), we use 
 in the following the equivalent 
parameterization
\begin{equation}
{\cal L}_{Y} = 
- g_\tau
\left(\cos\phi_\tau\bar{\tau}\tau + \sin\phi_\tau\bar{\tau}i\gamma_{5}\tau\right)
h \,,
\label{YukLa-phi}
\end{equation}
where $g_\tau$ is the effective strength of the $\tau$-Yukawa interaction and
$\phi_\tau$ describes the degree of mixing of the scalar and pseudoscalar 
 Higgs component which couple to $\tau$ leptons.
\begin{equation}
\label{Yuparphi}
g_\tau 
=
(\sqrt{2}G_{F})^{1/2}m_{\tau}\sqrt{a_{\tau}^{2}+b_{\tau}^{2}}
\, , 
\qquad
\tan\phi_\tau=\frac{b_{\tau}}{a_{\tau}}
\, .
\end{equation}

As to the mixing angle $\phi_\tau$, we remark the following. It
 is in general not universal, but specific to the $\tau$-Yukawa interaction. The reduced Yukawa couplings $a_f$, $b_f$ 
 to quarks and leptons $f$ are model-dependent. As an example one may
 consider  type-II two-Higgs doublet extensions of the Standard Model, where
 the ${\rm SU(2)}$ Higgs doublet $\Phi_2$ is coupled to the
 right-chiral $u$-type quarks and  the other doublet $\Phi_1$ is coupled to
 right-chiral $d$-type quarks and charged leptons. Referring to the
 model described for instance in \cite{Bernreuther:1992dz} one obtains
  in this case that $a_f$
 and likewise $b_f$ are identical for $d$-type quarks and charged
 leptons, while they differ in general for $u$-type quarks. Defining
 $\tan\phi_t=b_t/a_t$, where $a_t, b_t$ are the reduced Yukawa
 couplings of the top quark, one gets $\tan\phi_\tau=
\tan\alpha  \tan\beta \tan\phi_t$, 
where $\alpha$ denotes the mixing angle of the two CP-even neutral
components of the two Higgs doublet fields and 
$\tan\beta=v_2/v_1$ is the ratio
 of the vacuum expectation values of the Higgs doublets $\Phi_2$ and
 $\Phi_1$. For notational simplicity we  call $\phi_\tau$ the Higgs mixing
 angle.

For the SM Higgs boson, which is CP-even, one has $g_{\tau}=(\sqrt{2}G_{F})^{1/2}m_{\tau}$
and $\phi_\tau=0$. The ATLAS and CMS results on the 125
GeV resonance $h$ exclude that it is a pure pseudoscalar. In the following we investigate
how precisely a possible pseudoscalar component of $h$, i.e. $\sin\phi_\tau\neq0$,
can be determined in its $\tau$ decays at the LHC by means of the observables defined below.

The observables that we use \cite{Bernreuther:1993df,Bernreuther:1997af,Berge:2008wi,Berge:2008dr,Berge:2011ij,Berge:2013jra}
are based on $\tau$-spin correlations. The charged prongs, i.e., the
charged lepton $l=e,\mu$ in (\ref{taulept}), the charged pion in
(\ref{taupi}) - (\ref{taua1}), and the $a_{1}^{L,T}$ serve as $\tau$-spin
analyzers. The normalized distributions of polarized $\tau$ decays
to a $\pi^{\mp}$, a charged lepton $l=e,\mu$, a charged $\rho$
or $a_{1}$, and to a charged pion via $\rho$ and $a_{1}$ decay
are, in the $\tau$ rest frame, of the form 
\begin{eqnarray}
{\Gamma_{a}}^{-1}\mbox{d}\Gamma\left(\tau^{\mp}(\hat{{\bf
      s}}^{\mp})\to a^{\mp}(q^{\mp})+X\right) & = &
n\left(E_{\mp}\right)\left[1\pm b\left(E_{\mp}\right)\,\hat{{\bf
      s}}^{\mp}\cdot\hat{{\bf
      q}}^{\mp}\right]dE_{\mp}\frac{d\Omega_{\mp}}{4\pi} \, . \label{eq:dGamma_dEdOmega}
\end{eqnarray}
Here, ${\bf \hat{s}}^{\mp}$ denote the normalized spin vectors of
the $\tau^{\mp}$ and $E\mp$ and $\hat{{\bf q}}^{\mp}$ are the energies
and directions of flight of $a^{\mp}=l^{\mp},\pi^{\mp}$ in the respective
$\tau$ rest frame. The spectral functions $n$ and $b$ are given
in \cite{Berge:2011ij}. The function $b(E_{\mp})$ encodes the $\tau$-spin
analyzing power of the particle $a^{\mp}$. The $\tau$-spin analyzing
power is maximal for the direct decays to pions, $\tau^{\mp}\to\pi^{\mp}$,
and for $\tau^{\mp}\to a_{1}^{L,T,\mp}.$ (We recall that the $\tau$-spin
analyzing power of $a_{1}^{L-}$ and $a_{1}^{T-}$ is $+1$ and $-1$,
respectively.) For the other decays, the $\tau$-spin analyzing power
of $l^{\mp}$ and $\pi^{\mp}$ depends on the energy of these particles.
It can be optimized by judiciously chosen energy cuts.

\section{Distributions for Higgs production and decay to $\tau\tau$ at the
LHC\label{sec:dsigmaH_LHC}}

The hadronic differential cross section $d\sigma$ for Higgs production
at the LHC is given as a convolution of parton distribution functions
and the partonic differential cross section $d\hat{\sigma}_{ij}$
for the production of $h$ by partons $i$ and $j$ and subsequent
$h$ decay. For the decays \eqref{phitaudec} of $h$ to $\tau$ leptons
$d\hat{\sigma}_{ij}$ factorizes into a product of the squared $h$
production and decay matrix elements, as long as one neglects higher
order electroweak corrections that connect the production and $\tau$-decay
stage of $h$. The 125 GeV resonance $h$ is narrow, $\Gamma_{h}<4.2\Gamma_{h}^{SM}$
at 95$\%$ CL \cite{CMS:2014ala} where $\Gamma_{h}^{SM}=4.29$ MeV (see,
for instance \cite{Dittmaier:2011ti}). Therefore we can use the narrow width approximation
for $h$. In the following we are interested in the angular correlations
of the charged prongs $a^{+}$ and $a'^{-}$ in the decays \eqref{phitaudec}.
The characteristic features of these correlations depend only on the
$CP$ nature of $h$, but not on the details of its production. Therefore
we exhibit these correlations for the case of inclusive Higgs
production $i\, j\to h+X \to \tau^{-}\tau^{+} +X \to a^{-}a'^{+}+X$ (which
is dominated by gluon fusion). The structure of these correlations
applies also to other processes, for instance $h\,+\,{\rm jet}$ production
or $h$ production by vector boson fusion.

We choose a right-handed coordinate frame where the $\tau^{-}$ direction
of flight ${\bf \hat{k}}$ in the $\tau^{-}\tau^{+}$ zero-momentum
frame ($\tau\tau$ ZMF) defines the $z$ axis. The $\tau^{\pm}$ rest
frames are connected with the $\tau\tau$ ZMF by rotation-free Lorentz
boosts. In the formula \eqref{eq:dLHCsigma} below, 
$\theta_{\mp}=\angle({\bf \hat{k}},{\bf \hat{q}}_{\mp})$ are the polar
angles of  $a^{-}$ and $a'^{+}$, where ${\bf \hat{q}}_{\mp}$
are the directions of flight of $a^{-}$ and $a'^{+}$ in the $\tau^{\mp}$
rest frame, respectively, and 
\begin{equation}
\varphi=\phi_{-}-\phi_{+}\,,\qquad0\le\varphi\le2\pi\,,\label{eq:Define_varphi}
\end{equation}
is the difference of their azimuthal angles. With \eqref{YukLa-phi}
and \eqref{eq:dGamma_dEdOmega} we obtain for the differential partonic
cross section at leading order: 
\begin{eqnarray}
d\hat{\sigma}_{ij} & = & \frac{g_{\tau}^{2}m_{h}}{128\pi^{3}s\Gamma_{h}}\overline{\sum}\left|M\left(i\, j\to h\right)\right|^{2}\mbox{Br}_{_{\tau^{-}\to a^{-}}}\mbox{Br}_{_{\tau^{+}\to a'^{+}}}\label{eq:dLHCsigma}\\[1ex]
 &  & \times d\Omega_{\tau}dE_{-}d\Omega_{-}dE_{+}d\Omega_{+}n\left(E_{+}\right)n\left(E_{-}\right)\nonumber \\[1ex]
 &  & \times\left[1+b\left(E_{+}\right)b\left(E_{-}\right)\left(\cos\theta_{+}\cos\theta_{-}-\sin\theta_{+}\sin\theta_{-}\cos(\varphi-2\phi_\tau)\right)\right]\,.\nonumber 
\end{eqnarray}
Here, $\sqrt{s}$ is the partonic center-of-mass energy, $m_{h}$
is the mass of $h$, and $\phi_\tau$ is the Higgs mixing angle defined
in \eqref{YukLa-phi}. Moreover, we have put in \eqref{eq:dLHCsigma}
the $\tau$ velocity $\beta_{\tau}$ in the $h$ rest-frame equal
to one.

The angular correlations in \eqref{eq:dLHCsigma} reflect the $\tau$
spin correlations induced in the decay of $h$. Integrating Eq.~\eqref{eq:dLHCsigma}
with respect to $d\Omega_{\tau}d\cos\theta_{-}d\cos\theta_{+}$, the
differential partonic cross section takes the form 
\begin{eqnarray}
d\hat{\sigma}_{ij} & = & \frac{g_{\tau}^{2}m_{h}}{8\pi^{2}s\Gamma_{h}}\overline{\sum}\left|M\left(i\, j\to h\right)\right|^{2}\mbox{Br}_{_{\tau^{-}\to a^{-}}}\mbox{Br}_{_{\tau^{+}\to a'^{+}}}\label{eq:dLHCsigma2}\\[1ex]
 &  & \times d\varphi~dE_{-}dE_{+}n\left(E_{+}\right)n\left(E_{-}\right)\nonumber \\[1ex]
 &  & \times\left[1-b\left(E_{+}\right)b\left(E_{-}\right)\frac{\pi^{2}}{16}\cos(\varphi-2\phi_\tau)\right]\,.\nonumber 
\end{eqnarray}
Also this distribution encodes the CP nature of $h$. It was shown
in \cite{Berge:2008dr,Berge:2013jra} that the difference $\varphi$
of the azimuthal angles, which is equal to the angle between the signed
normal vectors of the $\tau^{-}\to a^{-}$ and $\tau\to a'^{+}$ decay
planes, can actually be measured in the zero-momentum frame of the
charged prongs $a^{-}$ and $a'^{+}$. This has the big advantage
that the $\tau^{\mp}$ momenta need not be reconstructed in experiments.

\section{Observables}
\label{Observables} 

Our method to determine the CP nature of a spin-zero
resonance $h$ in its decays \eqref{phitaudec} has been described
in detail in \cite{Berge:2008dr,Berge:2011ij,Berge:2013jra}. We recall
here its salient features. It requires the measurement of the 4-momenta
of the charged prongs $a^{-}$, $a'^{+}$ and their impact parameter
vectors  ${\bf {n}}_{\mp}$ in the laboratory
frame. The corresponding unit vectors are denoted by ${\bf
  \hat{n}}_{\mp}$.  The 4-vectors $n_{\mp}^{\mu}=(0,{\bf \hat{n}}_{\mp})$
are boosted into the $a^{-}a'^{+}$ ZMF and the spatial parts of the
resulting 4-vectors $n_{\mp}^{*\mu}$ are decomposed into their normalized
components ${\hat{n}}_{\textbar\textbar}^{*\mp}$ and ${\hat{n}}_{\perp}^{*\mp}$
which are parallel and perpendicular to the respective  3-momentum of $a^{-}$ and
$a'^{+}$. With the `unsigned' normal vectors ${\hat{n}}_{\textbar\textbar}^{*\mp}$
one determines the `unsigned' angle $\varphi^{*}$ between the $\tau^{-}\to a^{-}$
and $\tau\to a'^{+}$ decay planes in the $a^{-}a'^{+}$ ZMF: 
\begin{equation}
\varphi^{*}=\arccos({\bf \hat{n}}_{\perp}^{*+}\cdot{\bf
  \hat{n}}_{\perp}^{*-})\, , \qquad 0\leq\varphi^{*}\leq\pi\,.\label{phistar}
\end{equation}
The simultaneous measurement of (\ref{phistar}) and of the $CP$-odd
and $T$-odd triple correlation 
\begin{equation}
{\cal O}_{CP}^{*}={\bf \hat{p}}_{-}^{*}\cdot({\bf \hat{n}}_{\perp}^{*+}\times{\bf \hat{n}}_{\perp}^{*-})\,,\label{CP-oddTrip}
\end{equation}
where ${\bf \hat{p}}_{-}^{*}$ is the normalized $a'^{-}$ momentum
in the $a^{-}a'^{+}$ ZMF, allows to determine a `signed' angle (in
the range 0 to $2\pi$) between the $\tau^{-}\to a^{-}$ and $\tau\to a'^{+}$
decay planes in the $a^{-}a'^{+}$ ZMF, which is denoted by $\varphi_{CP}^{*}$,
by the following prescription: 
\begin{equation}
\varphi_{CP}^{*}=\left\{ \begin{array}{ccc}
\varphi^{*} & if & {\cal O}_{CP}^{*}\geq0\,,\\
2\pi-\varphi^{*} & if & {\cal O}_{CP}^{*}<0\,.
\end{array}\right.\label{phistar_CP}
\end{equation}
The distribution of \eqref{phistar_CP} is given by \eqref{eq:dLHCsigma2}
with $\varphi\to\varphi_{CP}^{*}$. In terms of this angle, the triple
correlation \eqref{CP-oddTrip} is given by $\sin\varphi_{CP}^{*}$.

\begin{figure}[t]
\includegraphics[height=6.4cm]{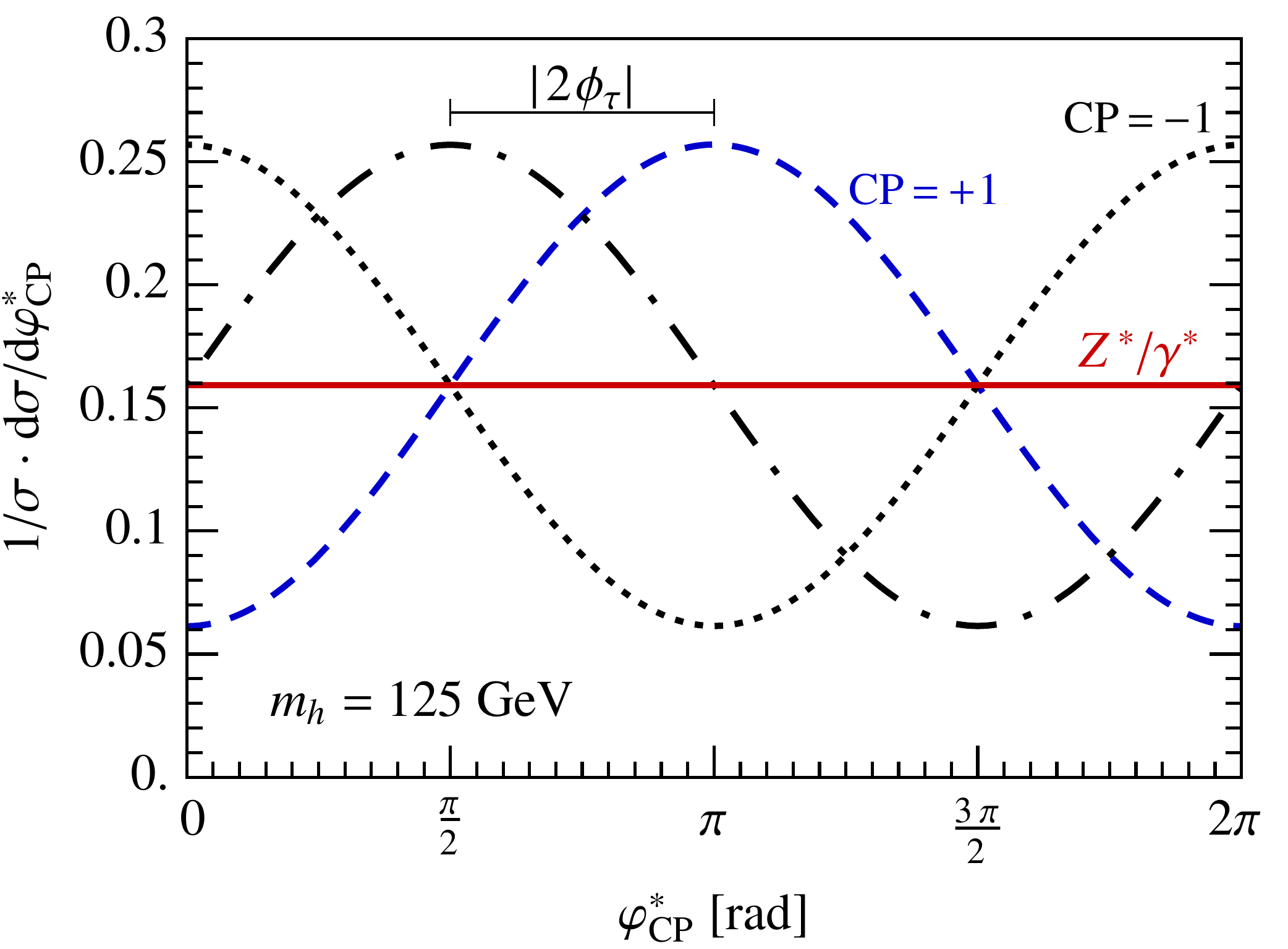}
\caption{Normalized $\varphi_{CP}^{*}$ distribution for the case where both
$\tau^{-}$ and $\tau^{+}$ decay to $\pi\nu$. The blue dashed line
is the distribution for a $CP$-even Higgs boson ($\phi_\tau=0$) and the
black long-dash dotted line corresponds to a $CP$ mixture with
$\phi_\tau=-\frac{\pi}{4}$. In addition, the distribution for a $CP$-odd
Higgs boson ($\phi_\tau=\pm{\pi}/{2}$, black dotted line) is also shown. The
solid red line is the distribution due to the $Z^*/\gamma^*\to\tau^{+}\tau^{-}$
background, cf. Sec.~\ref{sec:dsigmaZ_LHC}. }
\label{fig:h_pipi_detcuts} 
\end{figure}

The distribution of \eqref{phistar_CP} allows for an unambiguous
determination of the CP nature of $h$, that is, of the Higgs mixing
angle $\phi_\tau$. For illustration, the distribution of $\varphi_{CP}^{*}$
is shown in Fig.~\ref{fig:h_pipi_detcuts} for the decay mode $\tau^{-}\tau^{+}\to\pi^{+}\pi^{-}+2\nu$
for a CP-even and CP-odd Higgs boson and a  CP mixture.

By fitting the function $f=u\cos\left(\varphi_{CP}^{*}-2\phi_\tau\right)+{\rm v}$
to the measured differential distributions of $\varphi_{CP}^{*}$
for the respective final states $aa'$, one can determine the mixing
angle $\phi_\tau$. The function $f$ is subject to the constraint 
 $\int_{0}^{2\pi}d\varphi_{CP}^{*}f\,=\,2\pi{\rm v}\,=\,\sigma_{aa'}$,
where $\sigma_{aa'}$ is the $h$-production cross section including
the respective decay branching fractions. For a certain final state
$aa'$ the estimate of the statistical uncertainty of $\varphi_{CP}^{*}$
depends on the values of the associated parameters $u$ and ${\rm v}$.
Because $u$ and ${\rm v}$ describe the unnormalized distributions,
it is more convenient to compare the different channels with a normalized
quantity. The following asymmetry turns out to suit this purpose \cite{Berge:2013jra}:
\begin{eqnarray}
A^{aa'}\; & =\frac{1}{\sigma_{aa'}}\,\int_{0}^{2\pi}\!\! d\varphi_{CP}^{*}\left\{ d\sigma_{aa'}(u\cos(\varphi_{CP}^{*}-2\phi_\tau)>0)-d\sigma_{aa'}(u\cos(\varphi_{CP}^{*}-2\phi_\tau)<0)\right\} \nonumber \\
 & \,=\,\displaystyle{\frac{-4u}{2\pi{\rm v}}  }\,.\label{phiCP_asym}
\end{eqnarray}
This asymmetry can also be computed from \eqref{eq:dLHCsigma}, \eqref{eq:dLHCsigma2}.
In the absence of cuts, one obtains 
\begin{equation}
A^{aa'}=\frac{\pi}{8}\frac{\int dE_{a'^{+}}dE_{a^{-}}n\left(E_{a'^{+}}\right)n\left(E_{a^{-}}\right)b\left(E_{a'^{+}}\right)b\left(E_{a^{-}}\right)}{\int dE_{a'^{+}}dE_{a^{-}}n\left(E_{a'^{+}}\right)n\left(E_{a^{-}}\right)}\,.\label{phiCP_asym2}
\end{equation}
Eqs.~\eqref{phiCP_asym}, \eqref{phiCP_asym2} show that the values
of $A^{aa'}$ are independent of the mixing angle $\phi_\tau$ but do depend
on the product of the $\tau$-spin analyzing powers of $a$ and $a'$.
The larger $A^{aa'}$ the smaller the statistical error $\Delta\phi_\tau$
in this decay channel, for a given number of events. The $\tau$-spin
analyzing power, and thus $A^{aa'}$, is maximal for the direct decays
$\tau^{\mp}\to\pi^{\mp}$ and for $\tau^{\mp}\to a_{1}^{L,T\mp}.$
The $\tau$-spin analyzing power of the charged lepton in $\tau^{\mp}\to l^{\mp}$
and of the charged pion from $\tau^{\mp}\to\rho^{\mp}$ and $\tau^{\mp}\to a_{1}^{\mp}$
can be enhanced by applying an appropriate cut on the energy of the
lepton and the pion, respectively \cite{Berge:2011ij,Berge:2013jra}.

The background from Drell-Yan processes to Higgs production at the
LHC affects the respective distribution $d\sigma/d\varphi_{CP}^{*}$.
As will be shown below, this background contribution to the
$\varphi_{CP}^{*}$ distribution
 is flat for all charged prongs $a,a'$ if integrated over the full
phase space of the final states from the $\tau^\mp$ decays.
  Of course, this contribution decreases
the height of the normalized distribution and thus the magnitude of
the asymmetry \eqref{phiCP_asym}. In the next sections we investigate
which cuts may be used to significantly suppress this background.

\section{Drell-Yan production of $\tau^{-}\tau^{+}$\label{sec:dsigmaZ_LHC}}

Background reactions to the $h\to\tau^{-}\tau^{+}$ signal include
production of QCD multijets, $t\bar{t}$, single top, $W$ + jets,
$Z^{*}/\gamma^{*}$ + jets, $WW$, $WZ$, and $ZZ$. The Drell-Yan
process $Z^{*}/\gamma^{*}\to\tau^{-}\tau^{+}$ is an essentially
irreducible background to Higgs production by the reaction \eqref{lhcggjet}.
Because the mass of $h$ is relatively close to the mass of the $Z$
boson, an appropriate cut on the tau-pair invariant mass
$M_{\tau\tau}>M_{\rm cut}$
suppresses the photon contribution, but not the contributions from
$Z$ and the $Z\gamma$ interference term to the squared Drell-Yan
matrix element.

Because our method of determining the CP nature of $h$ uses the distribution
\eqref{eq:dLHCsigma2} in the $a^{-}a'^{+}$ ZMF, we need the corresponding
distribution for Drell-Yan production of $\tau$ pairs. The spin correlations
of the $\tau$ pairs produced by the intermediate vector bosons and
the subsequent angular correlations between $a^{-}$ and $a'^{+}$
differ from the correlations \eqref{eq:dLHCsigma} induced
by $h$ decay. In particular, unlike in $h$ decay%
\footnote{If one takes into account higher order electroweak corrections in
$h\to\tau\tau$,  longitudinal $\tau^\mp$ polarizations are  also induced
in $\tau$ pair production by $h$ decay \cite{Bernreuther:1997af},
which are however too small to be of relevance here.%
}, the $\tau^{\mp}$ samples are longitudinally polarized to some degree
due to the parity-violating couplings of the $Z$ boson. In order
to exhibit these features we consider Drell-Yan production of $\tau$
pairs and their subsequent decays to charged prongs $a^{-}$ and $a'^{+}$
to lowest order in the SM couplings. The corresponding parton reaction
is 
\begin{equation}
q\,+{\bar{q}}\,\to\gamma^{*},Z^{*}\to\tau^{-}\,+\,\tau^{+}\to a^{-}\,+\, a'^{+}\,+\, X \, .
 \label{DYtau0}
\end{equation}
For the partonic differential cross section which is analogous to
\eqref{eq:dLHCsigma} we obtain%
\footnote{We use the matrix elements given in \cite{Bernreuther:1993nd}, adapted
to the reactions \eqref{DYtau0}.%
} with \eqref{eq:dGamma_dEdOmega}, neglecting terms of order $m_{\tau}/\sqrt{s}$:
\begin{eqnarray}
d{\hat{\sigma}}_{DY}^{(0)}=\frac{1}{576\pi^{3}}\mbox{Br}_{_{\tau^{-}\to a^{-}}}\mbox{Br}_{_{\tau^{+}\to a'^{+}}}d\cos\theta_{-}d\cos\theta_{+}d\phi_{-}d\phi_{+}dE_{-}dE_{+}F(E_{i},\theta_{i},\phi_{i})\,,\label{dsDY0}
\end{eqnarray}
where $i=\pm$ and 
\begin{eqnarray}
F= & n(E_{-})n(E_{+})\sum\limits _{B_{1},B_{2}=Z,\gamma}a(B_{1},B_{2})\nonumber \\
 & \times\Bigg\{ v_{\tau}^{B_{1}}v_{\tau}^{B_{2}}\left[1-b(E_{+})b(E_{-})\left(\cos\theta_{+}\,\cos\theta_{-}+\frac{1}{2}\sin\theta_{+}\sin\theta_{-}\cos(\phi_{+}+\phi_{-})\right)\right]\nonumber \\
 & +\;\;\; a_{\tau}^{B_{1}}a_{\tau}^{B_{2}}\left[1-b(E_{+})b(E_{-})\left(\cos\theta_{+}\,\cos\theta_{-}-\frac{1}{2}\sin\theta_{+}\sin\theta_{-}\cos(\phi_{+}+\phi_{-})\right)\right]\nonumber \\
 & +\left(a_{\tau}^{B_{1}}v_{\tau}^{B_{2}}+a_{\tau}^{B_{2}}v_{\tau}^{B_{1}}\right)\left(b(E_{+})\cos\theta_{+}-b(E_{-})\cos\theta_{-}\right)\Bigg\}\,.\label{dsDY01}
\end{eqnarray}
The angles $\theta_{\pm}$, $\phi_{\pm}$ are the polar and azimuthal angles
of the $a^{-}$ and $a'^{+}$ in the $\tau\tau$ ZMF, where the direction
of the $\tau^{-}$
momentum is chosen to be the $z$-axis, and the momentum of the initial quark 
is located in the $x,z$-plane. Furthermore,
 \begin{eqnarray}
v_{f}^{\gamma}=Q_{f}e\quad(e>0)\,, & \qquad a_{f}^{\gamma}=0\,,\label{def21}\\
v_{f}^{Z}=e\ \frac{T_{3f}-2Q_{f}s_{\theta_{W}}^{2}}{2s_{\theta_{W}}c_{\theta_{W}}}\,, & \qquad a_{f}^{Z}=e\ \frac{T_{3f}}{2s_{\theta_{W}}c_{\theta_{W}}}\,,\label{def22}
\end{eqnarray}
and 
\begin{equation}
a(B_{1},B_{2})=s\ \frac{v_{q}^{B_{1}}v_{q}^{B_{2}}+a_{q}^{B_{1}}a_{q}^{B_{2}}}{D(B_{1})D^{*}(B_{2})}\,,\qquad D(B)=s-m_{B}^{2}+im_{B}\Gamma_{B}\,.\label{def32}
\end{equation}
Eq.~\eqref{dsDY01} shows that the angular correlations, which are
characteristic for the $\tau$ spin correlations induced by an intermediate
 spin-1 boson with vector and axial vector couplings, differ from those in \eqref{eq:dLHCsigma}.
The last term in \eqref{dsDY01} signifies the polarization of the
$\tau^{\pm}$ samples. Substituting $\phi_{-}=\varphi+\phi_{+}$ in
\eqref{dsDY01}, where $\varphi$ is defined in Eq.~\eqref{eq:Define_varphi},
and integrating \eqref{dsDY0} with respect to $\phi_{+}$ from $0$
to $2\pi$, the terms proportional to $\cos(\phi_{+}+\phi_{-})$ in
\eqref{dsDY01} vanish. That is, the resulting hadronic distribution
$d{\sigma}_{DY}^{(0)}/dE_{+}dE_{-}d\cos\theta_{+}d\cos\theta_{-}d\varphi$
is independent of $\varphi$ for any final state $a^{-}a'^{+}$. This
is displayed, for $Z^{*}/\gamma^{*}\to\tau\tau\to\pi\pi$, in Fig.~\ref{fig:h_pipi_detcuts}
by the solid red line.

We find it instructive to investigate this feature in more detail. As a
result we obtain a proposal for calibrating the distribution of
$\varphi$ respectively $\varphi_{CP}^{*}$ with Drell-Yan $\tau\tau$
events, see below. For definiteness, we choose the charged prongs
$a^-,a^+ = \pi^-,\pi^+$ in the following discussion.
 Eq.~\eqref{dsDY01}  shows that the $Z^{*}/\gamma^{*}$
contribution to the $\varphi$ distribution 
is flat only if \eqref{dsDY01} is integrated over the full $2\pi$
range of $\phi_{+}$ (or alternatively of  $\phi_{-}$).
 The  $\varphi$ distribution will deviate from a flat line if the phase space
 of one of the pions is restricted.
 For instance, if one demands the $\pi^{-}$ momentum to lie in  the plane defined by
$\phi_{-}=0$, the contribution of, for instance, the   pure photon exchange to the
$\varphi$ distribution (which follows from the first line in the
 curly bracket  of Eq.~\eqref{dsDY01}) is proportional to $1-c\cos\varphi$. This
 distribution and, therefore, the distribution of   $\varphi_{CP}^{*}$ differs from a
 flat line.

How can this be probed experimentally? We define  a variable
$\cos{\tilde\alpha}_-$ by  
\begin{eqnarray}\label{defcath}
\cos{\tilde\alpha}_{-} & = & \left|\frac{{\bf \hat{e}_{z}}\times{\bf \hat{k}}_{L-}}{\left|{\bf \hat{e}_{z}}\times{\bf \hat{k}}_{L-}\right|}\cdot\frac{{\bf \hat{p}}_{L-}\times{\bf \hat{k}}_{L-}}{\left|{\bf \hat{p}}_{L-}\times{\bf \hat{k}}_{L-}\right|}\right|
\end{eqnarray}
\vspace{1mm}

which allows to classify  the $Z^{*}/\gamma^{*}\to\tau^-\tau^+\to\pi^-\pi^+$
 events into events where the $\pi^-$ is  `nearly coplanar' and `nearly perpendicular'
 to the $q\tau$ production plane in the laboratory frame. In
 \eqref{defcath} the unit vectors ${\bf \hat{k}}_{L-}$ and
 ${\bf \hat{p}}_{L-}$ are the $\tau^-$ and  $\pi^{-}$ directions of
 flight in the laboratory frame and ${\bf \hat{e}}_{z}$ points along
 the direction of one of the proton beams. The range of 
 ${\tilde\alpha}_-$ is  $0\le{\tilde\alpha}_{-}\le {\pi}/{2}$.
 Events with $\pi^{-}$ being  `nearly coplanar' (`nearly
 perpendicular') are defined by 
 demanding ${\tilde\alpha}_{-}<\pi/4$ (${\tilde\alpha}_{-}>\pi/4$). In order to define
 a  discriminating variable in terms of measurable quantities, we
 use the impact parameter vector ${\bf \hat{n}_{-}}$
 (cf. Sec.~\ref{Observables}) instead of ${\bf \hat{k}}_{L-}$ and
 replace \eqref{defcath} by 

\vspace{-5mm}
 
\begin{eqnarray}
\cos\alpha_{-} & = &
\left|\frac{{\bf\hat{e}_{z}}\times{\bf\hat{p}}_{L-}}{\left|{\bf\hat{e}_{z}}\times{\bf\hat{p}}_{L-}\right|}
\cdot\frac{{\bf \hat{n}_{-}}\times{\bf \hat{p}}_{L-}}{\left|{\bf \hat{n}_{-}}\times{\bf \hat{p}}_{L-}\right|}\right|\,\,.\label{cosalphabeta}
\end{eqnarray}
\vspace{1mm}
This variable is nearly identical to  \eqref{defcath}, i.e., events with $\pi^{-}$ being  `nearly coplanar' (`nearly
 perpendicular') are in the following defined by   requiring ${\alpha}_{-}<\pi/4$ (${\alpha}_{-}>\pi/4$).

\begin{figure}[t]
\includegraphics[height=5.3cm]{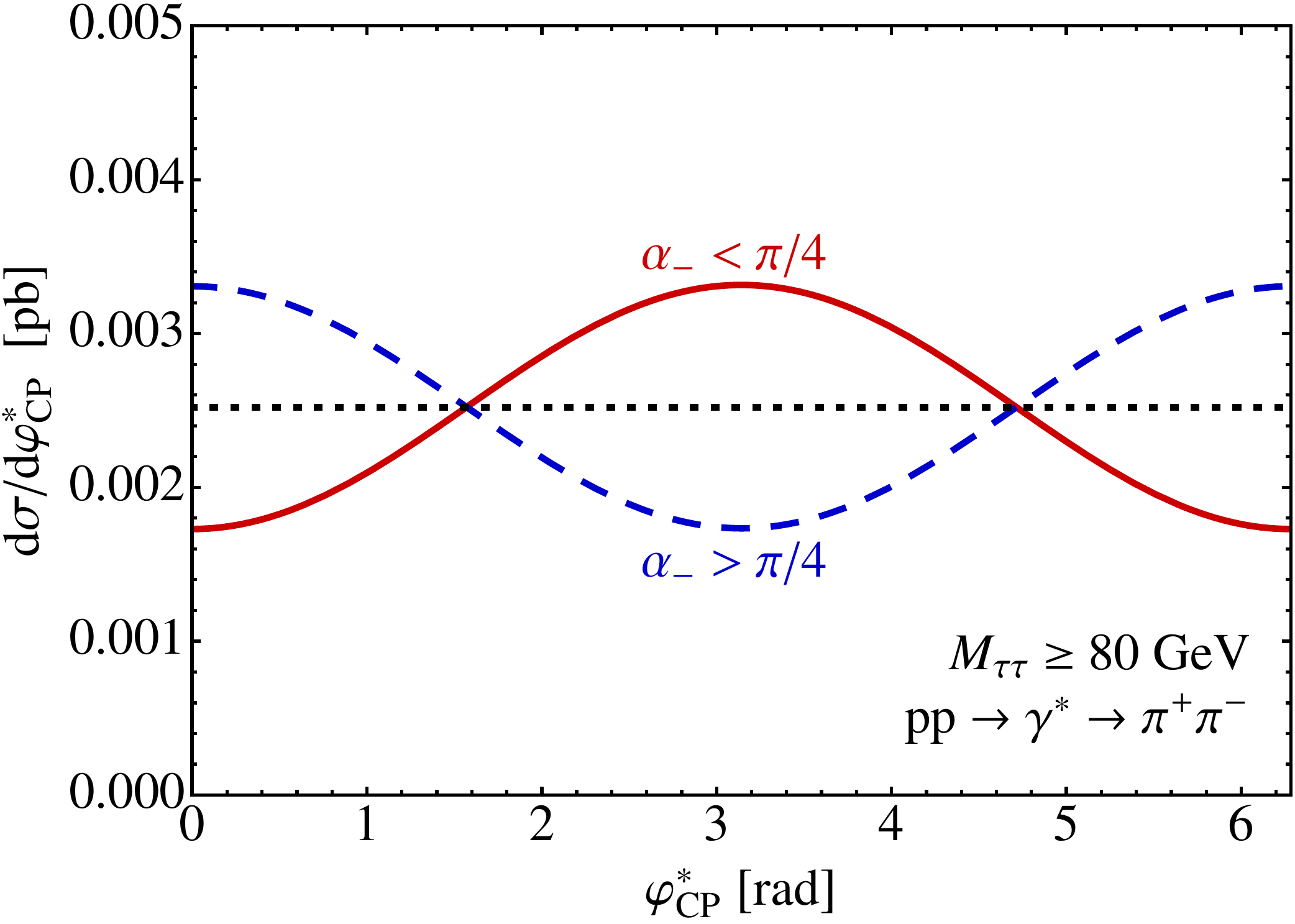}
\hspace{12mm}\includegraphics[height=5.3cm]{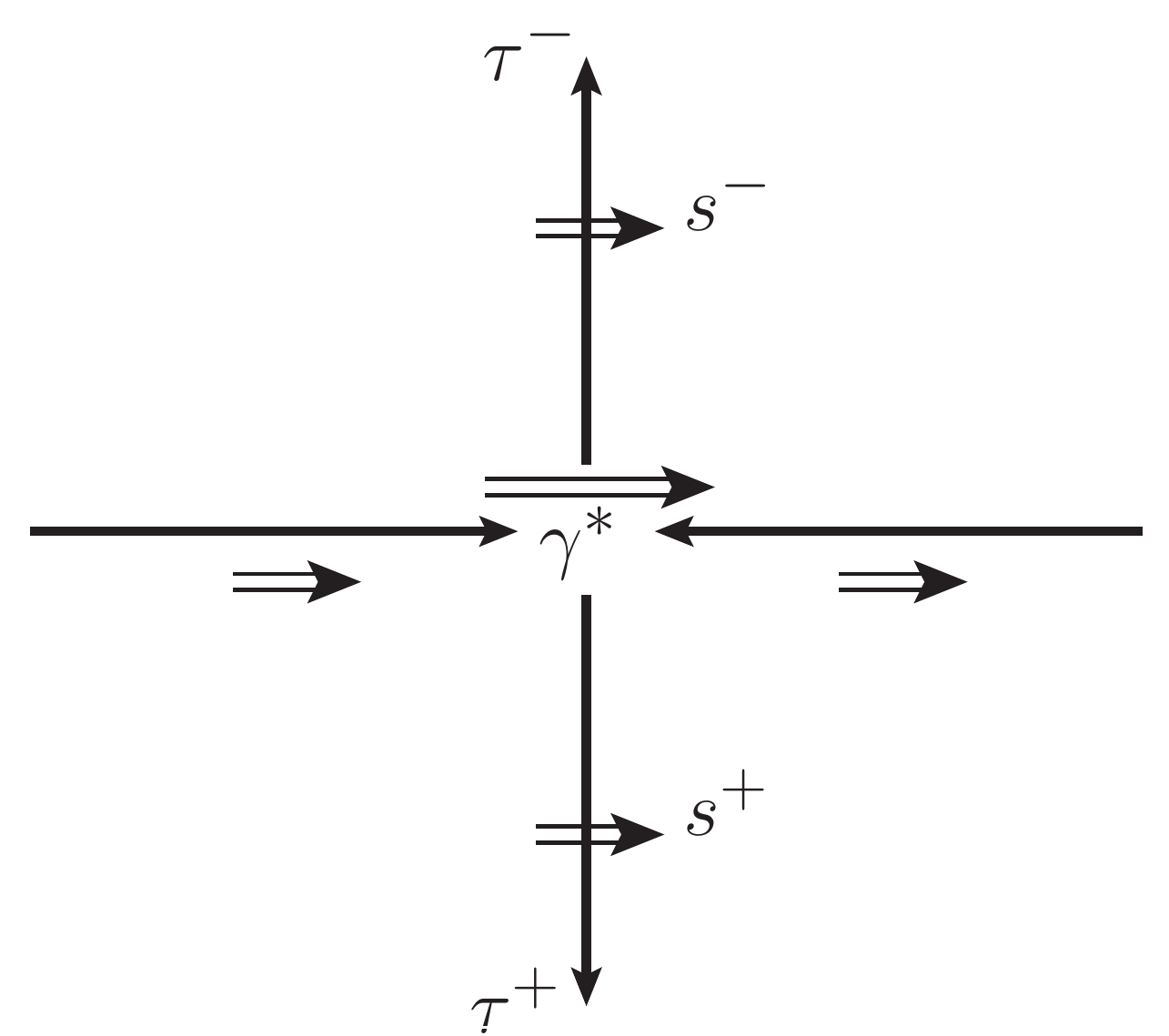}
\caption{Drell-Yan production of $\tau^-\tau^+$ and subsequent
 decays to pions  via  photon exchange, $pp\to\gamma^{*}\to\tau^{-}\tau^{+}\to\pi^{+}\pi^{-}+2\nu$.
Left: The un-normalized distribution of 
 $\varphi_{CP}^{*}$ for events with  $\pi^{-}$ `nearly coplanar' ($\alpha_{-}<\pi/4$,
red solid line) and events   with $\pi^{-}$ `nearly perpendicular' ($\alpha_{-}>\pi/4$, dashed blue
line) to the  $q\tau$ production plane. The dotted black line is half
of the sum of  the two distributions. The cuts
$M_{\tau\tau}\ge80$~GeV and $|\eta_{\pi^{\pm}}|\le1$ were used.
Right: Spin configuration for $q{\bar
  q}\to\gamma^{*}\to\tau^{-}\tau^{+}$ events where the $\tau^-\tau^+$
are produced orthogonal to the beam direction.}
\label{fig:LHC_Phi_CP_star_nonflat_gamma} 
\end{figure}

Let us first consider Drell-Yan production of $\tau^-\tau^+$ and subsequent
 decays to pions  via  photon exchange,
 $pp\to\gamma^{*}\to\tau^{-}\tau^{+}\to\pi^{+}\pi^{-}+2\nu$.
 The  $\varphi_{CP}^{*}$ distributions (computed with the first line in
 the curly bracket of \eqref{dsDY01}) are
 shown in Fig.~\ref{fig:LHC_Phi_CP_star_nonflat_gamma}, left plot,
 for events with  $\pi^{-}$ `nearly coplanar' ($\alpha_{-}<\pi/4$,
red solid line) and events   with $\pi^{-}$ `nearly perpendicular' ($\alpha_{-}>\pi/4$, dashed blue
line) to the  $q\tau$ production plane.

The distribution for events  with  $\pi^{-}$ `nearly coplanar' is
enhanced for  $\varphi_{CP}^{*}\sim\pi$ which corresponds to 
 $\pi^{-}$  and $\pi^{+}$ being (nearly) antiparallel. In order to
 understand this let us consider $q{\bar q}\to\gamma^*\to \tau^-\tau^+$
 where, for illustration, the $\tau$ pair is emitted perpendicular to
 the incoming quark direction. The  $\tau$ pair is produced in a
 s-wave and the resulting spin configuration for this type of events
 is shown in  the right plot of Fig.~\ref{fig:LHC_Phi_CP_star_nonflat_gamma}, i.e., the 
  $\tau^{-}$ and $\tau^{+}$ spin projections onto the quark axis are
  parallel. The $\tau^\pm$ decay distributions
  \eqref{eq:dGamma_dEdOmega} tell us that the $\pi^-$ ($\pi^+$) are
  then preferentially emitted in (opposite to)
   the direction of the  $\tau^-$  ($\tau^+$)  spin, which means
  that the  $\varphi_{CP}^{*}$ distribution is enhanced for $\phi_{-}
  - \phi_{+}\sim \pi$.

 On the other hand if the $\pi^{-}$ is emitted `nearly perpendicular'  to
 the $q\tau$ production
plane ($\alpha_{-}>\pi/4$), the $\varphi_{CP}^{*}$ distribution
 is enhanced  at $\varphi_{CP}^{*}\sim 0$ and $2\pi$,
 cf. the left plot of   Fig.~\ref{fig:LHC_Phi_CP_star_nonflat_gamma}. Again this can
 be understood from the right plot of of Fig.~\ref{fig:LHC_Phi_CP_star_nonflat_gamma}
  and the $\tau^\pm$ decay distributions
  \eqref{eq:dGamma_dEdOmega}. The projection of the spin of $\gamma^*$
  and thus the projection of the total $\tau^-\tau^+$ spin onto the
  axis orthogonal to the $q\tau$ production plane is zero, i.e.,
 the $\tau^-$ and $\tau^+$ spins are predominantly
 anticorrelated with respect to this axis (`up-down' and `down-up').
 Therefore,  the momenta of  the $\pi^{-}$ and $\pi^{+}$ are
 preferentially parallel in this case. 
 The $\varphi_{CP}^{*}$ distributions for the two sets of
 events ($\alpha_{-}<\pi/4$ and $\alpha_{-}>\pi/4$) add up exactly to
 a flat line as stated above and already shown in Fig.~\ref{fig:h_pipi_detcuts}. 

 Let us now consider \eqref{DYtau0} with the intermediate $Z$ boson. In view of the
 analysis of Sec.~\ref{sec:Numerical-Results-LHC} below, we apply a cut
 on the $\tau$-pair invariant mass $M_{\tau\tau}\geq M_{\rm cut}$ ($M_{\rm cut} \gtrsim 80$ GeV).
 Then  \eqref{DYtau0} is dominated by $Z$-boson exchange. The strengths of the
 vector and axial vector couplings of the $\tau$ leptons imply that the differential cross section
  \eqref{dsDY0} is dominated by the second line of the curly 
 bracket in \eqref{dsDY01}, i.e., by  $\tau$-pair production
  through the axial vector current, which corresponds to p-wave production of
 $\tau^-\tau^+$. The resulting $\tau$ spin correlations differ from those induced by $\gamma^*$
 exchange discussed above. In the case of axial vector
 production the $\tau^-$ and $\tau^+$ spin projections onto the quark
 axis are predominantly anticorrelated (excluding the forward and backward regions), while
 the  $\tau^-$ and $\tau^+$ spin projections onto the axis orthogonal to the
 $q\tau$ production plane are  predominantly correlated.  Therefore, in the case of axial
  vector production, the  $\varphi_{CP}^{*}$ distributions for events  
 with $\pi^{-}$ emitted    `nearly coplanar' and `nearly perpendicular', respectively,
 are opposite to the corresponding distributions for $\gamma^*$ exchange shown in
  Fig.~\ref{fig:LHC_Phi_CP_star_nonflat_gamma}.   

\begin{figure}[t]
\includegraphics[height=5.1cm]{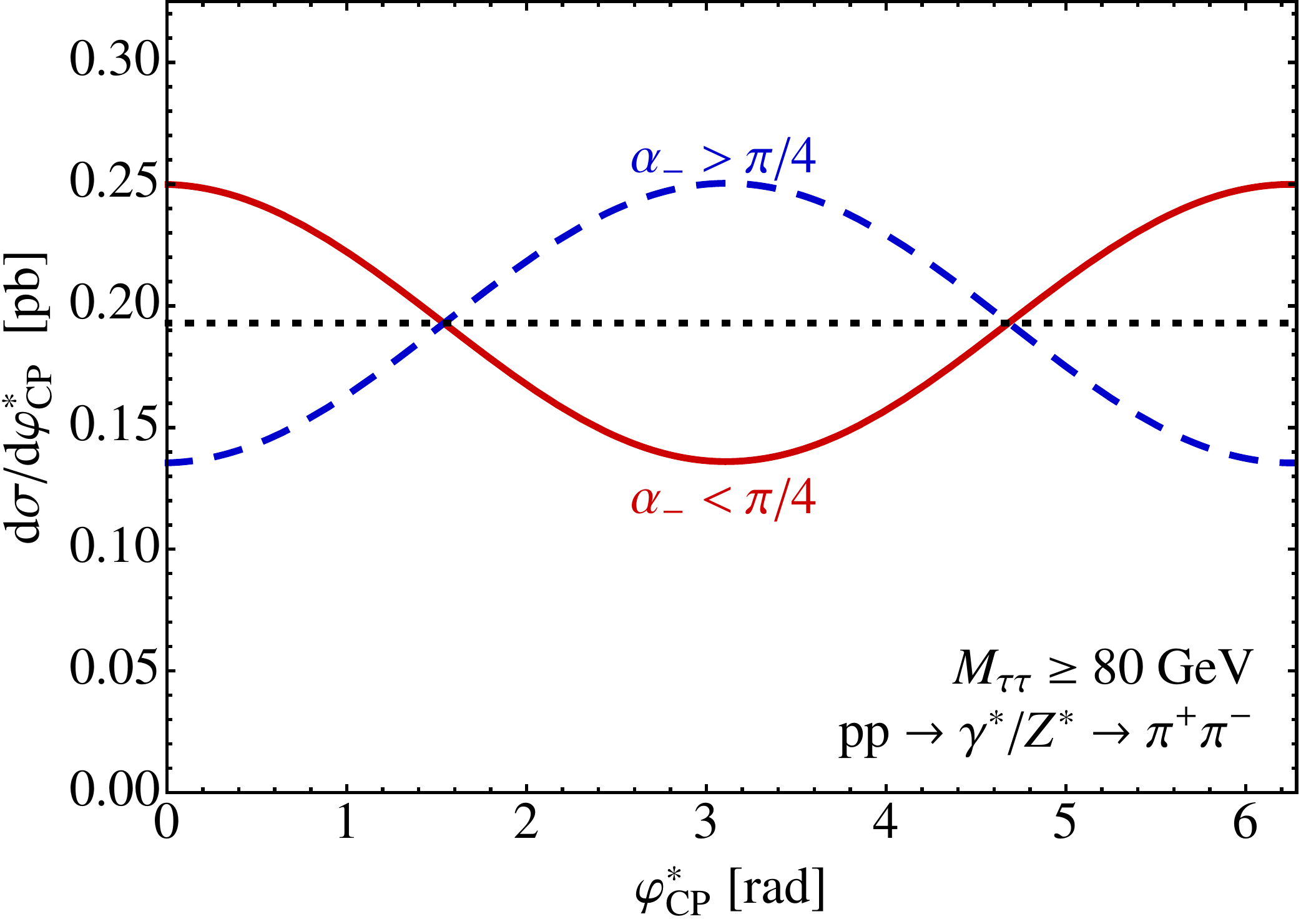}\hspace*{.5cm}
\includegraphics[height=5.1cm]{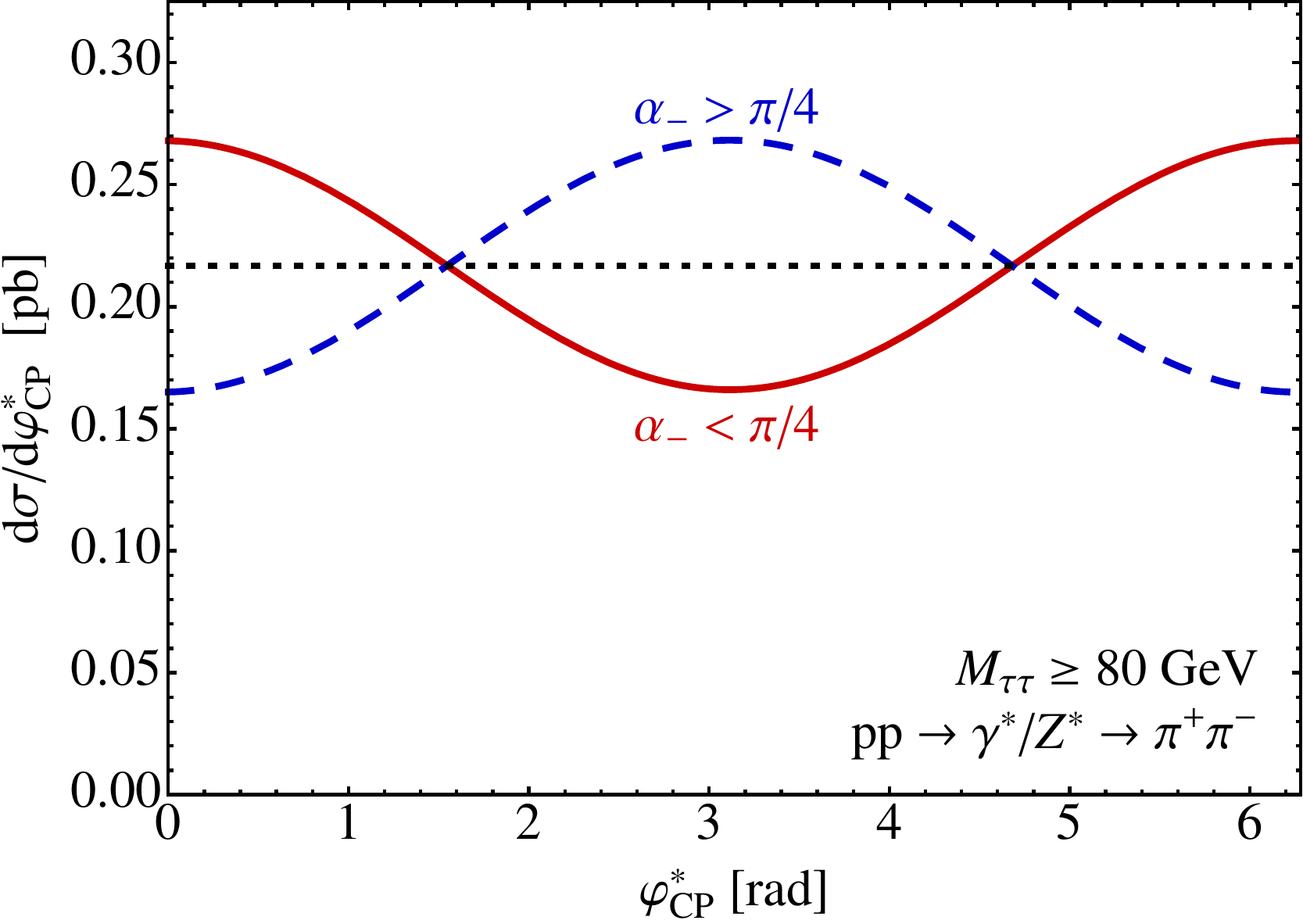}
\caption{Drell-Yan production of $\tau^-\tau^+$ and subsequent
 decays to pions via $Z^*/\gamma^*$  exchange, $pp\to Z^{*}/\gamma^{*}\to\tau^{-}\tau^{+}\to\pi^{+}\pi^{-}+2\nu$.
The  un-normalized distribution of 
 $\varphi_{CP}^{*}$ for events with  $\pi^{-}$ `nearly coplanar' ($\alpha_{-}<\pi/4$,
red solid line) and events   with $\pi^{-}$ `nearly perpendicular' ($\alpha_{-}>\pi/4$, dashed blue
line) to the  $q\tau$ production plane. The dotted black line is half
of the sum of  the two  distributions.
 The cuts  $M_{\tau\tau}\ge80$~GeV and $|\eta_{\pi^{\pm}}|\le1$ were
 used. Left plot, LO QCD. Right plot, NLO QCD.}
\label{fig:LHC_Phi_CP_star_nonflat} 
\end{figure}

The left plot of Fig.~\ref{fig:LHC_Phi_CP_star_nonflat} shows the corresponding
un-normalized $\varphi_{CP}^{*}$ distributions computed with
 the complete tree-level differential cross section   \eqref{dsDY0}, \eqref{dsDY01}.
 The cuts  $M_{\tau\tau}\ge80$~GeV and $|\eta_{\pi^{\pm}}|\le1$ on the  pseudo-rapidity of the pions
 were applied. The shape of these distributions reflect the outcome of the discussion made in the previous
 paragraph\footnote{As to the signal reaction $pp\to h\to
   \tau^-\tau^+\to \pi^-\pi^+2\nu$ we remark the following. For a
   Higgs boson of any CP nature, the un-normalized $\varphi_{CP}^{*}$
   distributions have the same shape for events with $\pi^-$ being
   nearly coplanar and nearly perpendicular.}.
                                Again, the two distributions add up to a flat line.

Contrary to the case of an intermediate Higgs boson, the tree-level distributions
\eqref{dsDY01} will be affected by higher-order QCD corrections to \eqref{DYtau0} because
of the correlations of the $\tau$ spins with the initial-state parton
momenta. We have computed the respective differential distributions
for 
\begin{equation}
pp\to Z^{*},\gamma^{*}+\, X\,\to\tau^{-}\,+\,\tau^{+}+\, X\,\to a^{-}\,+\, a'^{+}\,+\, X\,
 \label{DYNLOpp}
\end{equation}
at next-to-leading order (NLO) in the QCD coupling $\alpha_{s}$,
taking the $\tau$  correlations in the virtual and real corrections
into account. We calculated the respective $\tau$ spin density matrices
for $q{\bar{q}}\to Z^{*},\gamma^{*}(g)\to\tau^{-}\tau^{+}(g)$
and $gq({\bar{q}})\to Z^{*},\gamma^{*} q({\bar{q}})
\to\tau^{-}\tau^{+}q({\bar{q}})$
at order $\alpha_{s}$. The soft and collinear divergences are
treated with Catani-Seymour dipole subtraction \cite{Catani:1996vz}
with collinear factorization in the ${\overline{{\rm MS}}}$ scheme.

As to the NLO QCD $\varphi_{CP}^{*}$ distributions 
for $Z^{*}/\gamma^{*}\to  \tau^{-}\tau^{+} \to \pi^-\pi^+ + X$ for events
with $\alpha_{-}<\pi/4$  and $\alpha_{-}>\pi/4$: they are displayed in the
 right plot Fig.~\ref{fig:LHC_Phi_CP_star_nonflat}. The comparison with
 the LO distributions shows that  the order $\alpha_s$ QCD
 corrections amount to about $12\%$ 
and the shapes of these NLO distributions remain essentially the same as the LO
distributions.  

The results shown in Fig.~\ref{fig:LHC_Phi_CP_star_nonflat} suggest that one may use the
 Drell-Yan events $pp\to Z^*/\gamma^* \to \tau\tau \to \pi\pi\nu{\bar\nu}$, which are quite
 abundant at the LHC, for experimentally calibrating and validating the  $\varphi_{CP}^{*}$ distribution(s) before
 this observable is used to determine the CP nature of the Higgs boson $h$.

The NLO QCD  polar angle  and energy distributions of
the charged prongs $aa'$, which can be used for background suppression,
will be discussed in the next section.

\section{Numerical Results for the LHC}
\label{sec:Numerical-Results-LHC}

In this section we consider Higgs production~(\ref{lhcggjet}) and
decay into $\tau$ pairs~(\ref{phitaudec}) at the LHC for a collider
center-of-mass energy $\sqrt{S}=14$~TeV. We analyze how the differences
 between the Higgs-boson induced and $Z^*/\gamma^*$ induced 
 $\tau$ spin correlations, respectively the differences between the resulting
 $aa'$ angular and energy correlations can be used to reduce  the $Z^*/\gamma^*$
 background.   At the end of this section we estimate the precision
$\Delta\phi_\tau$ with which the Higgs mixing angle may be determined
in the $h\to\tau\tau$ decay mode at the LHC.

As long as no cuts on the final state particles/jets are applied,
the normalized distributions  \eqref{eq:dLHCsigma} for the signal reaction \eqref{lhcggjet}
hold also when higher order QCD corrections are taken into account.
 In order to obtain NLO QCD  distributions with cuts for \eqref{lhcggjet}, we compute 
 the two-dimensional  distributions of the Higgs boson transverse
 momentum and rapidity by means of the computer code
 MCFM \cite{Campbell:2010ff,Campbell:2006xx,Campbell:2010cz}. We include
 $h$ decay into $\tau$ pairs with $\tau$ spin correlations and their subsequent
 decays into charged prongs by Monte Carlo simulation.
 We put $m_{h}=125$ GeV and assume the
$gg\to h$ amplitude to be the same as in the SM, and we take $\Gamma_{h}=\Gamma_{h}^{SM}=4.29$~MeV.
  For estimating
signal-to-background ratios we take the K factor for inclusive Higgs production,
 ${\rm K}_{NNLO/NLO}=1.35$
into account, calculated with the computer code
  HNNLO \cite{Catani:2007vq,Grazzini:2008tf,Grazzini:2013mca}.
 We calculated the 
  angular distributions analogous to \eqref{dsDY0} and corresponding
  energy distributions for the background
reactions  \eqref{DYNLOpp} also to NLO QCD (cf. above). For the numerical evaluation
we used the parton distribution functions CT10 \cite{Lai:2010vv} with
$\alpha_{s}(m_{Z})=0.1180$ and two-loop running in $\alpha_{s}$.
Moreover, we used $\alpha(M_{Z})=1/128.89$ and the weak mixing angle
$\sin^{2}\theta_{W}=0.2228$. As a default value for the renormalization
and factorization scale $\mu$ we have set $\mu=\mu_{R}=\mu_{F}=m_{h}$.

The ATLAS and CMS experiments, which reported evidence for $h\to\tau\tau$
\cite{ATLtauconf,Chatrchyan:2014nva}, have reconstructed the $\tau\tau$
invariant mass with methods described in \cite{Elagin:2010aw} and
\cite{Chatrchyan:2014nva}, respectively. The Higgs-boson signal appears
as an enhancement of $d\sigma/dM_{\tau\tau}$ in a mass window around
$M_{\tau\tau}=m_{h}$ over the background which is mostly due to $Z^{*}/\gamma^{*}\to\tau\tau$.
If not stated otherwise, we apply in the following sections a cut
$M_{\tau\tau}>100$ GeV which strongly suppresses the background from
$Z^{*}/\gamma^{*}\to\tau\tau$.

\subsection{$\tau^{+}\tau^{-}$ spin correlations and subsequent polar
 angle and energy  correlations\label{susec:inklh}}

\begin{figure}[tb]
\includegraphics[height=6.3cm]{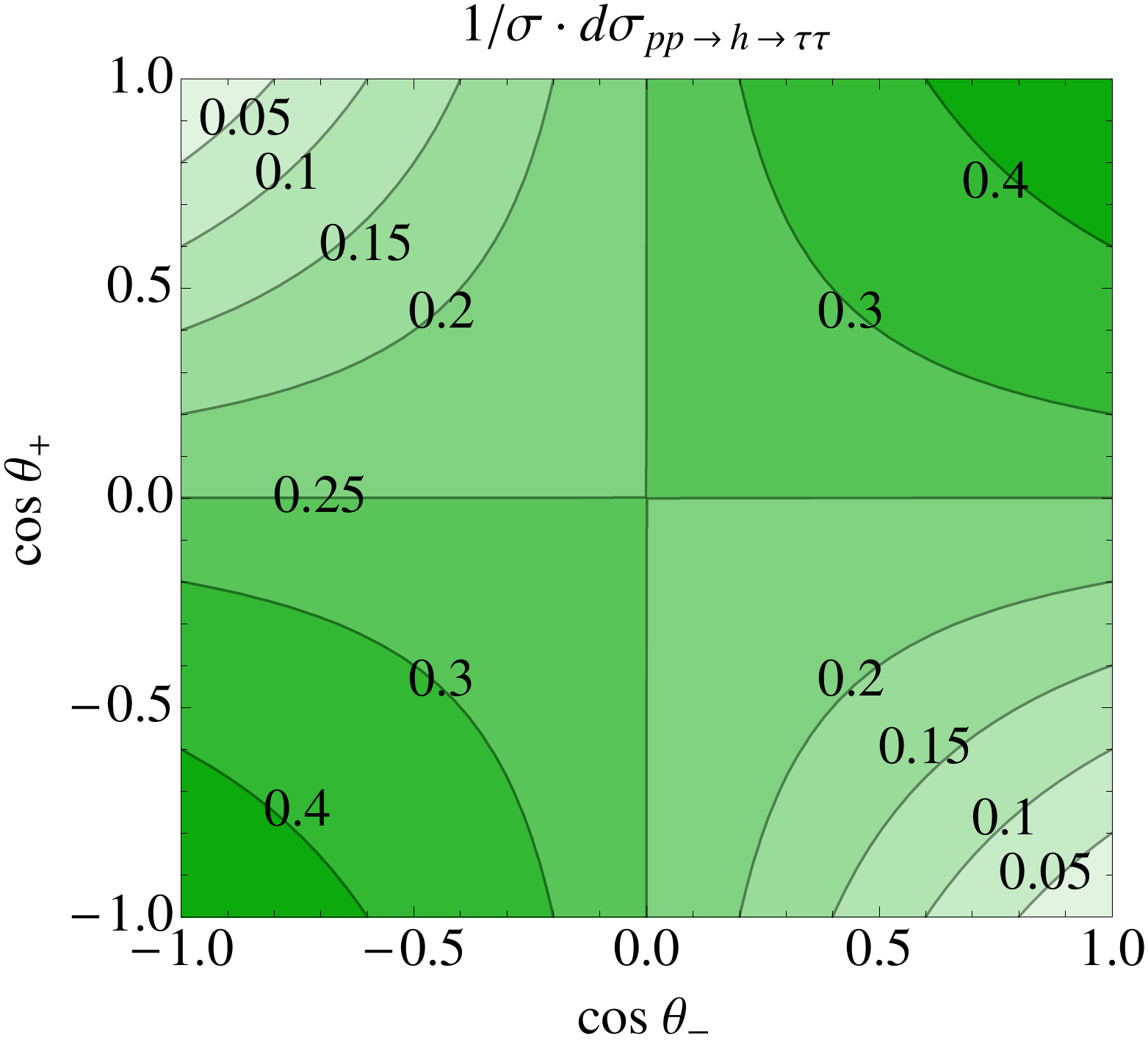}\hspace{13mm}
\includegraphics[height=6.3cm]{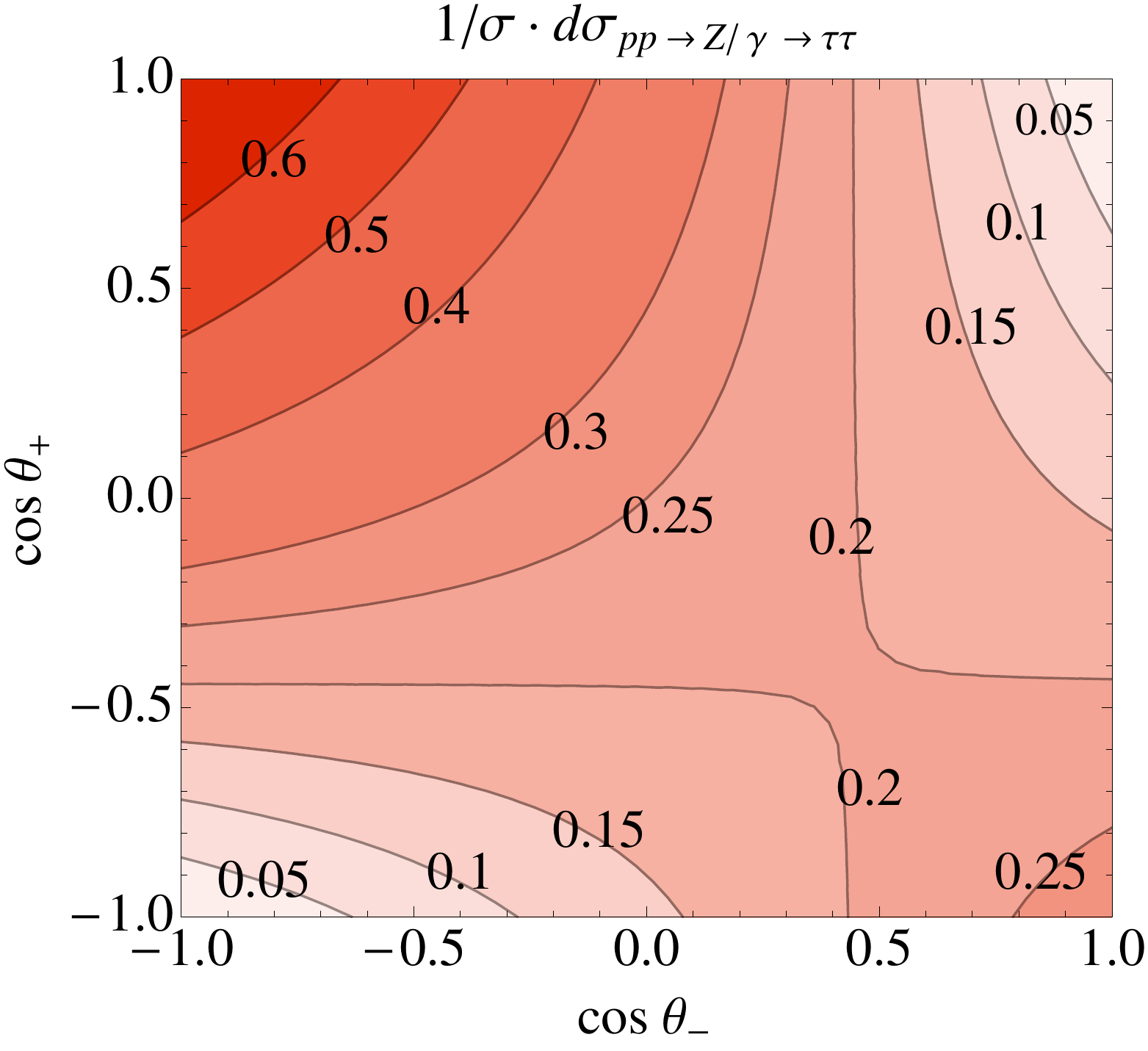}
\caption{LHC, $\sqrt{S}=14$~TeV, $M_{\tau\tau}\ge100$~GeV. Production of
$h+X$ (left) of arbitrary CP nature and of $Z^{*}/\gamma^{*}+X$ (right) with subsequent
decay of the respective boson to $\tau^{-}\tau^{+}\to\pi^{+}\pi^{-}+2\nu$.
The plots show the  distributions $\sigma^{-1}d\sigma/d\cos\theta_{-}d\cos\theta_{+}$
without cuts on the pion momenta. }
\label{fig:LHC_True_costheta_pm_distribution_2D} 
\end{figure}

We analyze the distributions $d\sigma/d\cos\theta_{-}d\cos\theta_{+}$
and associated energy distributions for the signal and background
reactions, where the helicity angles $\theta_{\mp}$ of $a^{-}$ and
$a'^{+}$ are defined as in Sec.~\ref{sec:dsigmaH_LHC}. In the
left and right plot of Fig.~\ref{fig:LHC_True_costheta_pm_distribution_2D}
 the normalized distributions $\sigma^{-1}d\sigma/d\cos\theta_{-}d\cos\theta_{+}$ at NLO QCD
are shown for the signal and background reaction for the $\tau^{-}\tau^{+}\to\pi^{-}\pi^{+}+2\nu$
decay mode. No cuts on the pion transverse momentum $p_T^\pi$  or
the pion rapidity are applied. Solid grey contour lines denote constant
values. The normalized signal distribution is given by $(1+\cos\theta_{-}\cos\theta_{+})/4$,
cf. Eq.~\eqref{eq:dLHCsigma}, and becomes maximal for $\cos\theta_{-}\cos\theta_{+}\to1$.
The background distribution contains terms proportional to $1-\cos\theta_{-}\cos\theta_{+}$
due to the $Z^{*}/\gamma^{*}$ induced $\tau$ spin correlations and,
in addition, terms linear in $\cos\theta_{\mp}$ which signify a longitudinal
$\tau^{\mp}$ polarization generated by the parity-violating $Z\tau\tau$
coupling. The background can be reduced by applying a cut on the product
of cosines, $\cos\theta_{-}\cos\theta_{+}>-c_{min}$ $(c_{min}>0)$,
where $c_{min}$ should be different for $\cos\theta_{-}\to\pm1$,
in view of the unsymmetric background distribution. These cuts should
not be too hard because that part of the signal cross section which
is sensitive to the Higgs $CP$ mixing angle $\phi_\tau$ is proportional
to $\sin\theta_{-}\sin\theta_{+}$, cf. Eq.~\eqref{eq:dLHCsigma}.

The uncertainties due to scale variations  of these NLO distributions are as follows.
 The normalized signal
distribution in Fig.~\ref{fig:LHC_True_costheta_pm_distribution_2D},
left, is identical to the normalized LO distribution because  production
and decay of the Higgs boson factorizes. In order to estimate  the scale
uncertainty of the normalized $Z^{*}/\gamma^{*}$ distribution in
Fig.~\ref{fig:LHC_True_costheta_pm_distribution_2D}, right, we vary
the scale $\mu$ between $\mu=m_{h}/2$ and $\mu=2m_{h}$ and calculate
the deviation $[\sigma^{-1} d\sigma(\mu=m_{h})-\sigma^{-1} d\sigma(\mu)]/ \sigma^{-1}d\sigma(\mu=m_{h})$
for each value of $\cos\theta_{-}$ and $\cos\theta_{+}$. Apart from a
small region in 
the lower left and upper right corner of
Fig.~\ref{fig:LHC_True_costheta_pm_distribution_2D}, right,
 the maximal deviation of each point from the respective value of $1/\sigma\cdot
d\sigma(\mu=m_{h})$ is small, about $\pm3\%$.

\begin{figure}[t]
\includegraphics[height=6.3cm]{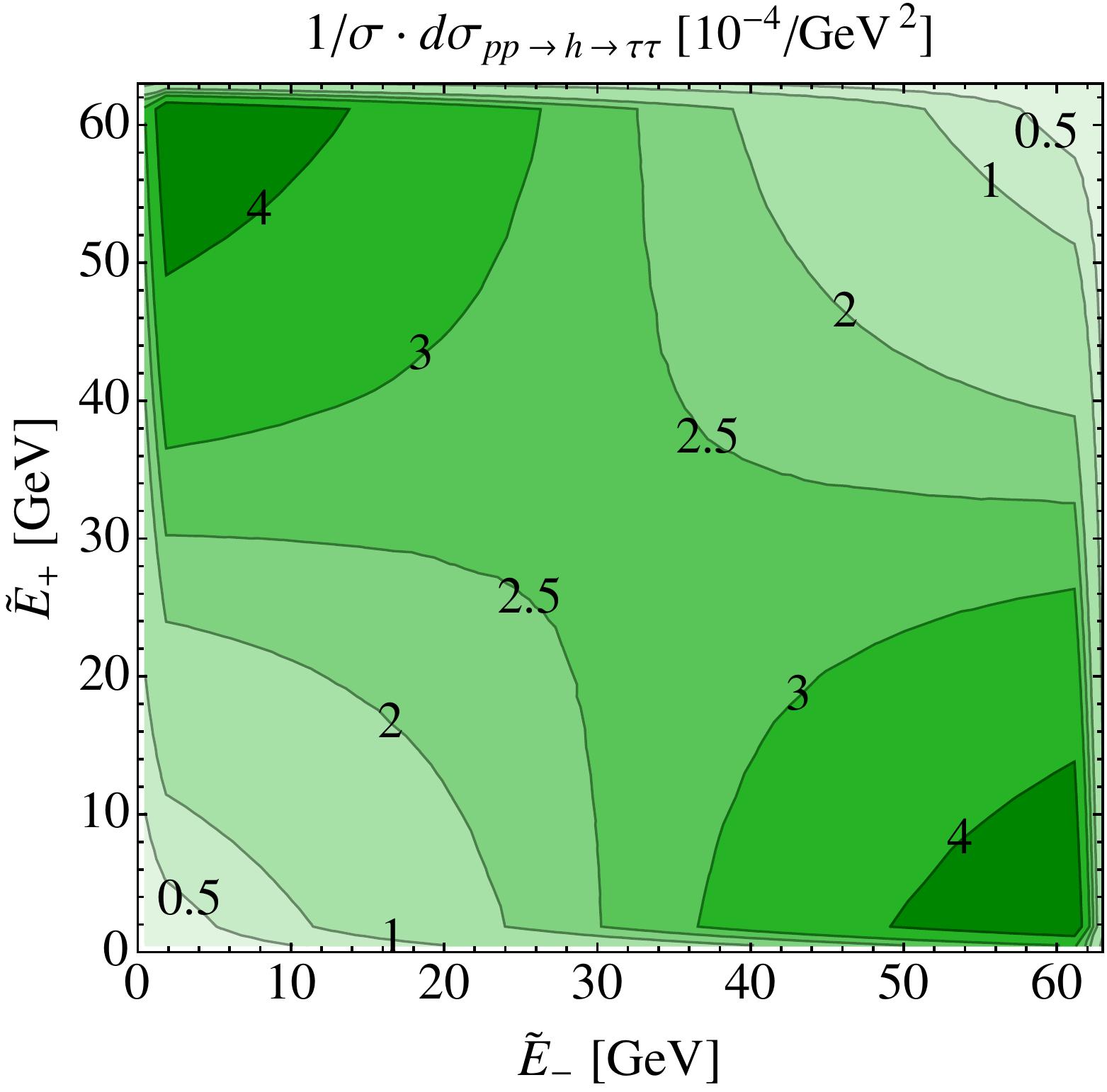}\hspace{13mm}
\includegraphics[height=6.3cm]{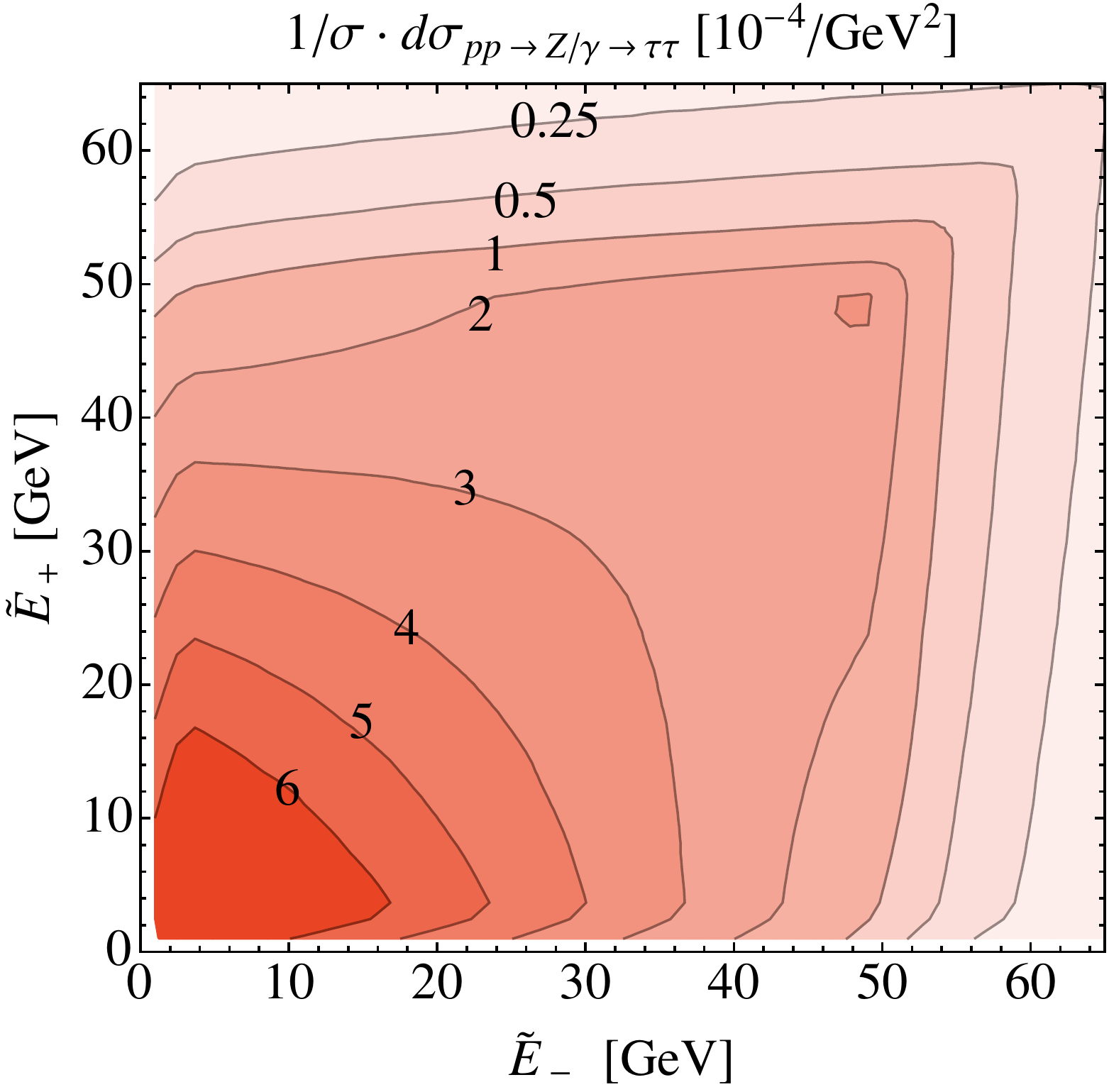}
\caption{LHC, $\sqrt{S}=14$~TeV, $M_{\tau\tau}\ge100$~GeV. Production of
$h+X$ (left) of arbitrary CP nature and of $Z^{*}/\gamma^{*}+X$ (right) with subsequent
decay of the respective boson to $\tau^{-}\tau^{+}\to\pi^{+}\pi^{-}+2\nu$.
The plots show the  distributions $\sigma^{-1}d\sigma/d{\tilde{E}}_{-}d{\tilde{E}}_{+}$
without cuts on the pion momenta. }
\label{fig:LHC_True_Epm-inkl_distribution_2D} 
\end{figure}

Both distributions in Fig.~\ref{fig:LHC_True_costheta_pm_distribution_2D}
are affected if a $p_{T}$ cut on
the pion momenta is applied. A $p_T^\pi$  cut mostly removes events
in the vicinity of $\cos\theta_{-}=-1$ and $\cos\theta_{+}=+1$,
because in these phase-space regions the pion is emitted opposite
to the corresponding $\tau$ direction of flight and, therefore, its
energy in the $\tau\tau$ ZMF is small.

Cuts on $\cos\theta_{\mp}$ may be unrealistic because, at the LHC,
the reconstruction of the $\tau$ rest frames  is complicated,
even for hadronic $\tau$ decays. This is because the partonic center-of-mass
energy is unknown for a certain event and the decay of each $\tau\tau$
pair involves at least two neutrinos. However, one can approximately
reconstruct the $\tau\tau$ ZMF with a  fitting procedure~\cite{Elagin:2010aw}
 and determine the pion energies in this frame, denoted
by ${\tilde{E}}_{\mp}$ in the following. The energies ${\tilde{E}}_{\mp}$
are related to $\cos\theta_{\mp}$ by boosts. In Fig.~\ref{fig:LHC_True_Epm-inkl_distribution_2D}
the NLO QCD distributions $\sigma^{-1}d\sigma/d{\tilde{E}}_{-}d{\tilde{E}}_{+}$
are displayed for the signal and background reaction. Solid grey contour
lines denote constant values. The normalized signal distribution in
Fig.~\ref{fig:LHC_True_Epm-inkl_distribution_2D}, left, shows that  $h\to \tau\tau$ events
decay preferably into one pion with a large energy and one pion with
a small energy in the $\tau\tau$ ZMF. On the other hand, the 
 right plot of Fig.~\ref{fig:LHC_True_Epm-inkl_distribution_2D} shows
 that  in the case of the $Z^{*}/\gamma^{*}\to \tau\tau$ background,
 events where both pion energies are small
   are strongly enhanced.

If one applies a cut on the transverse momenta of the $\pi^{\mp}$,
for instance, $p_T^\pi \ge20$ GeV, the number of events with small
${\tilde{E}}_{\mp}$ are reduced.  At LO QCD this cut
removes all events with ${\tilde{E}}_{\mp}<20$~GeV 
because  the transverse momentum of the $h,\, Z^{*},\,\gamma^{*}$
boson is zero. At NLO QCD this is lifted to some extent because the
finite transverse momentum of the respective boson results in 
$p_T^\pi \gtrsim 20$~GeV
even if ${\tilde{E}}_{\mp}<20$~GeV in the $\tau\tau$ ZMF.

The normalized distributions of Fig.~\ref{fig:LHC_True_Epm-inkl_distribution_2D}
suggest the application of  cuts on the energies ${\tilde{E}}_{\mp}$ in order to enhance
the signal-to-background ratio. This ratio is enhanced by  rejecting events where
both  ${\tilde{E}}_{+}$ and ${\tilde{E}}_{-}$ are smaller
than, for instance,  $20$~GeV, or where both energies are larger than $45$~GeV. However, one should
not  reject regions including  ${\tilde{E}}_{+}\sim{\tilde{E}}_{+}\sim30$~GeV which 
corresponds to $\sin\theta_{-}\sin\theta_{+}\sim1$. Here the sensitivity
to $\varphi_{CP}^{*}$ and therefore to the mixing angle $\phi_\tau$ is largest (cf. Eq.~(\ref{eq:dLHCsigma})).


\subsection{The distribution  of $\varphi_{CP}^{*}$ for $h\to\tau\tau$  \label{susec:phi_star_CP}}

\subsubsection{Direct $\tau^{+}\tau^{-}\to\pi^{+}+\pi^{-}+2\nu$ decay}
\label{susec:helicity_correllation}
\begin{figure}[t]
\includegraphics[height=5.2cm]{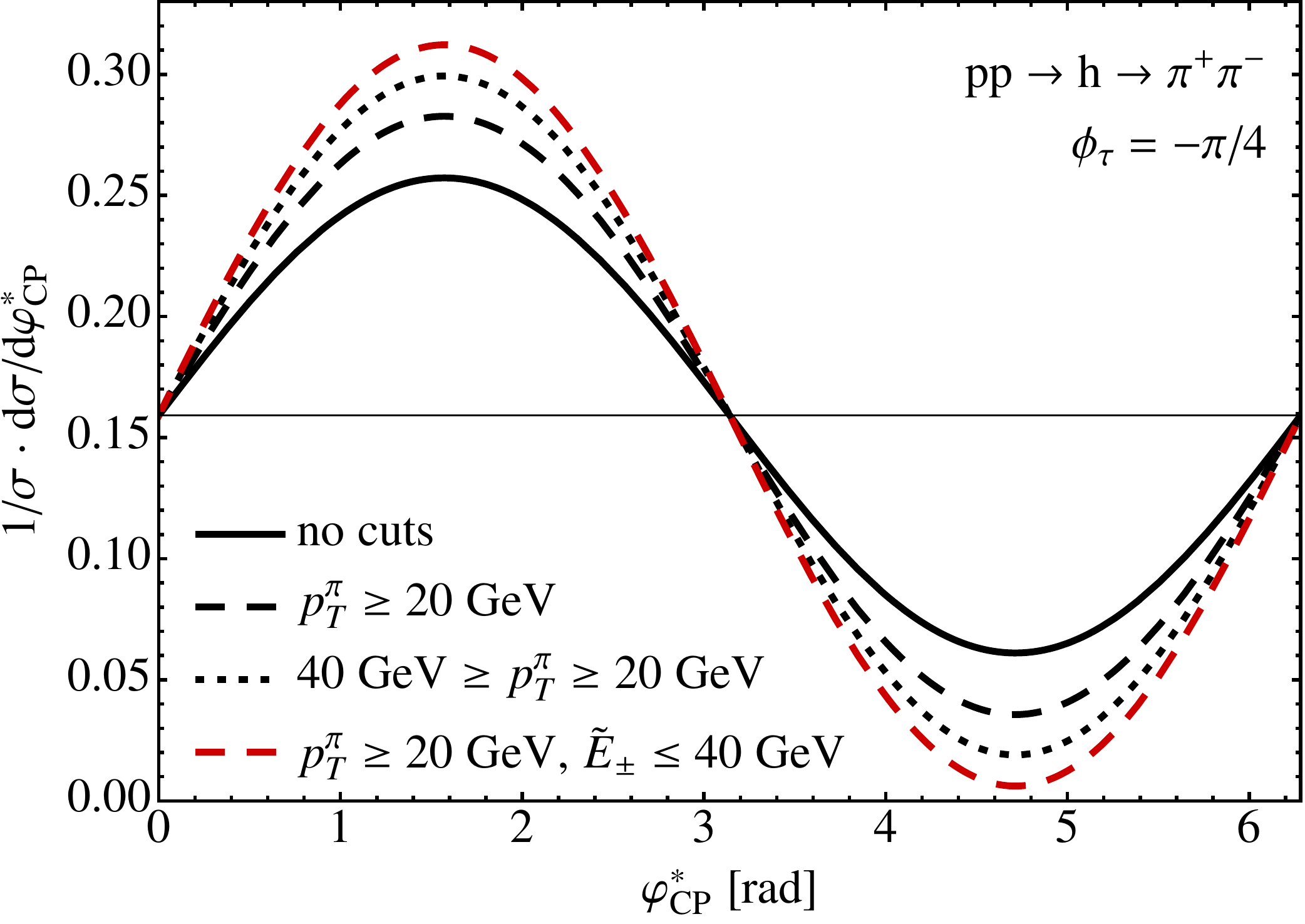}
\hspace{7mm}\includegraphics[height=5.2cm]{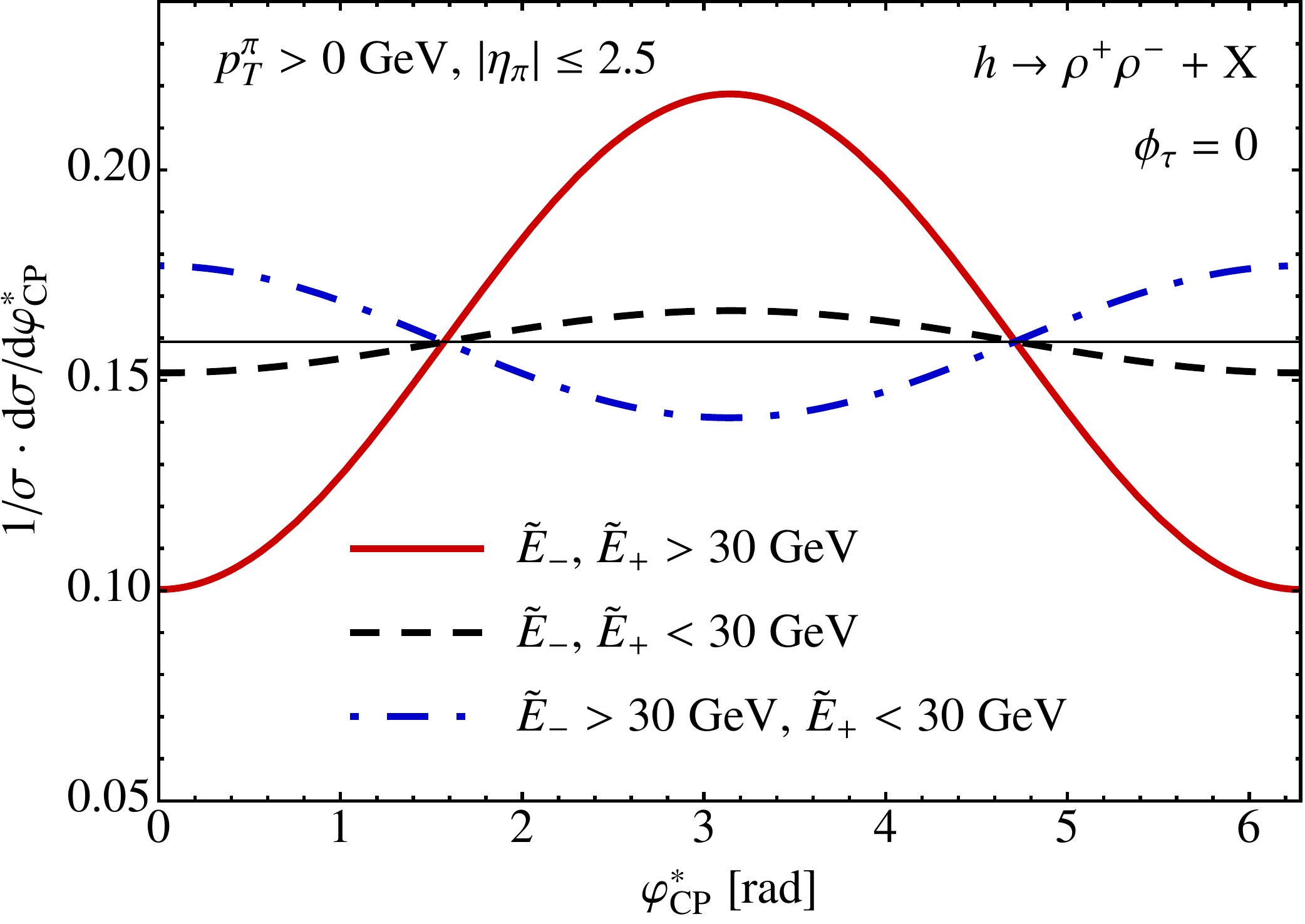}
\caption{
  Left: $pp\to h\to\tau^{-}\tau^{+}\to\pi^{+}\pi^{-}+2\nu$, where
   $h$ is  assumed to
 be a CP mixture with mixing angle  $\phi_\tau=-{\pi}/{4}$.
 Dependence of the distribution  $\sigma^{-1}d\sigma/d\varphi_{CP}^{*}$ on different
kinematical cuts. Right:  $h\to\tau^{+}\tau^{-}\to\rho^{+}\rho^{-}+2\nu$ for
 a CP-even Higgs boson.
 Dependence of the distribution  $\sigma^{-1}d\sigma/d\varphi_{CP}^{*}$ on cuts on the
 energies ${\tilde{E}}_{\mp}$ of the charged pions. The horizontal
 lines in both plots are drawn to guide the eye.}
\label{fig:LHC_phi_cut-dependence} 
\end{figure}

The normalized $\varphi_{CP}^{*}$ distribution  without cuts is shown in Fig.~\ref{fig:h_pipi_detcuts}
 for  $h\to\tau^{+}\tau^{-}\to\pi^{+}\pi^{-}+2\nu$.
 If no cuts on the final-state particles are applied, the distribution is the same also for other Higgs
production modes,  or if higher order QCD corrections
are included. The distribution will change if kinematical cuts like
$p_T^\pi$  cuts are applied, because the $\varphi$ distribution
  results from the term proportional
to   $\sin\theta_{+}\sin\theta_{-}$ in Eq.~\eqref{eq:dLHCsigma}.
Enhancing the region $\cos\theta_{+}\cos\theta_{-}\sim0$ of Fig.~\ref{fig:LHC_True_costheta_pm_distribution_2D},
left, increases the  asymmetry defined in Eq.~\eqref{phiCP_asym}.
The dependence of the normalized $\varphi_{CP}^{*}$ distribution  at NLO QCD  on cuts on 
$p_T^\pi$   and on ${\tilde{E}}_{\mp}$  is displayed
in Fig.~\ref{fig:LHC_phi_cut-dependence}, left, for a Higgs mixing angle
$\phi_\tau=-{\pi}/{4}$. The solid black line shows the distribution without cuts. The corresponding
asymmetry is $A^{\pi\pi}=39.3\%$. If a cut  $p_{T}^{\pi}\ge20$~GeV
is applied the asymmetry, associated with the distribution  shown by the dashed black line, 
 increases to $A_{NLO}^{\pi\pi}=49.5\%$ ($A_{LO}^{\pi\pi}=50.2\%$).
  For the Higgs-boson production mode~(\ref{lhcggjet}) 
   the  Higgs-boson  transverse momentum is, on average, small. Therefore the $p_{T}^{\pi}$ cut removes
 events  with $E_{\pm}^{\tau\tau}\lesssim20$~GeV in the distribution displayed in  Fig.~\ref{fig:LHC_True_Epm-inkl_distribution_2D}, left.
 For these events the value of
 $\sin\theta_\pm$ is small for at least
one of the pions. Therefore the  value of the product $\sin\theta_{+}\sin\theta_{-}$ is on average
 rather large for the remaining events. This is why the asymmetry $A$ is increased by this cut.
  If, in addition, events with large  transverse pion momenta are rejected by selecting, for instance,
  events with  $40{\rm \, GeV}\ge p_{T}^{\pi}\ge 20$~GeV, the asymmetry is further
enhanced to $56\%$ ($A_{LO}^{\pi\pi}=59.5\%$). The corresponding  $\varphi_{CP}^{*}$ distribution 
 is given  by the   dotted black line in Fig.~\ref{fig:LHC_phi_cut-dependence}.
 Cuts on the pion energies will also change this distribution and the
 resulting asymmetry. The  $\varphi_{CP}^{*}$ distribution shown by
   dashed red line in Fig.~\ref{fig:LHC_phi_cut-dependence}, left,
  results from applying the cuts $p_{T}^{\pi}\ge20$~GeV  and $\tilde{E}_{\pm}\le40$~GeV.
The corresponding  asymmetry is $61.2\%$ ($A_{LO}^{\pi\pi}=61.8\%$).

Notice that all these cuts do not change the location  of the maximum of
the $\varphi_{CP}^{*}$ distribution. Furthermore, cuts on the pseudo-rapidity
of the pions $\eta_{\pi}$ do not change the normalized $\varphi_{CP}^{*}$
distributions displayed  in Fig.~\ref{fig:LHC_phi_cut-dependence}.

\subsubsection{Other $\tau$ decay modes}
\label{suse:othertaum}

At the end of Sec.~\ref{susec:inklh}, cuts on the
 energies
 ${\tilde{E}}_{\mp}$ of the charged pions
 were suggested for the direct $\tau\to\pi$ decays, 
 in order to enhance the signal-to-background
ratio and the asymmetry~\eqref{phiCP_asym}. 
 If the Higgs mixing angle $\phi_\tau$ is to be determined from
  other $\tau$ decay
modes with the method described in Sec.~\ref{Observables}, e.g.
from  $\tau^{+}\tau^{-}\to\rho^{+}\rho^{-}+2\nu$ and subsequent $\rho^{\pm}\to\pi^{\pm}+\pi^{0}$
decay, these cuts on  ${\tilde{E}}_{\mp}$  can,
however, not be used for background suppression in this case. 
 This is because the $\tau$-spin analyzing power
of the charged pion from $\rho$ decay is energy-dependent, cf. for
instance Fig.~4 in \cite{Berge:2011ij}, where this analyzing power
is shown as a function of the pion energy in the $\tau$ rest frame.
This energy is related by a boost to the $\pi^{\pm}$ energy ${\tilde{E}}_{\pm}$
in the $\tau\tau$ ZMF. Dividing the ${\tilde{E}}_{-}$, ${\tilde{E}}_{+}$
phase space into four regions, two with ${\tilde{E}}_{\pm}\le30$~GeV,
${\tilde{E}}_{\pm}\ge30$~GeV and two, where one energy is smaller
than and the other one larger than $30$~GeV, we show 
in Fig.~\ref{fig:LHC_phi_cut-dependence}, right, the resulting normalized
distributions of the angle $\varphi_{CP}^{*}$. The distribution given by
       the red solid line,
which has the largest asymmetry~\eqref{phiCP_asym}, results from events where both energies
${\tilde{E}}_{\mp}\ge30$~GeV. For events with ${\tilde{E}}_{\mp}\le30$~GeV the distribution is almost flat (dashed black line). 
For events with
${\tilde{E}}_{-}\ge30$~GeV and ${\tilde{E}}_{+}\le30$~GeV (dot-dashed
blue line) the resulting asymmetry is also quite small.
Furthermore, the  $\varphi_{CP}^{*}$ distribution is shifted in this case
by  an angle $\pi$ with respect to the solid red line. This is because for
 $\tau\to\rho\to \pi$ decay, the function $b(E)$ in \eqref{eq:dGamma_dEdOmega}, which encodes the $\tau$-spin analyzing power
  of the charged pion for this decay mode, is negative   for ${\tilde{E}}\le30$~GeV.
  The  asymmetry~\eqref{phiCP_asym} is largest for
 events with  ${\tilde{E}}_{\mp}\ge30$~GeV
because if ${\tilde{E}}_{\pm}$ are large,   the pion energies in the respective
$\tau$ rest frames are also large on average. In this energy range
 the $\tau$-spin analyzing power of the charged pion from $\rho$
decay is large (and positive).

\subsubsection{Impact of measurement
  uncertainties\label{sub:Impact-of-measurement}}

The  normalized $\varphi_{CP}^{*}$ distributions are affected by
measurement uncertainties, in particular by the uncertainties
associated with the measurements of the directions  ${\bf \hat{n}}_{\mp}$
of the impact parameters  of the charged prongs $a^{-}$, $a'^{+}$
(cf. Sec.~\ref{Observables}). In order to assess the effect of these 
 uncertainties on the  distributions  $\varphi_{CP}^{*}$ for the
 various $\tau\tau$ decay modes with Monte Carlo methods, we 
 have ``smeared'' the relevant
quantities with a Gaussian distribution function $\propto \exp(-(X/{\sigma)}^{2}/2)$.
 Here $X$ denotes the generated quantity (coordinate in position
space, momentum component, energy) and $\sigma$ its expected standard deviation.

The primary vertex (PV), i.e., the Higgs-boson production/decay vertex is
varied   along and transverse to the beam axis 
with $\sigma_{z}^{PV}= 20\mu m$ and $\sigma_{tr}^{PV}=10\mu m$,
respectively. In the following, we discuss the effect of smearing in
some detail for the $\tau^-\tau^+\to\pi^-\pi^+$ decay mode.
 The intersection point of the impact parameter vector  ${\bf {n}}_{\mp}$ with the respective track of the charged pion
 $\pi^\mp$ 
 is varied by $\sigma_{tr}^{\pi}=10\mu m$ within a circle transverse
to the direction of the pion momentum. Furthermore we assume the angular
resolution of the charged $\pi^\mp$ track at its intersection point
with  ${\bf {n}}_{\mp}$  to be distributed with $\sigma_{\theta}^{\pi}=1$~mrad, and
 the resolution of the $\pi^{\pm}$ energy is taken to be  $\Delta E^{\pi}/E^{\pi}=5\%$.
 With these values, suggested in
\cite{Gennai:2006,Tarrade:2007zz}, we arrive at the conclusion
 that a rather precise measurement of the Higgs mixing angle 
 $\phi_\tau$ is possible at the LHC, see below.

First, we determine the average length $\langle|{\bf
  n}_\mp|\rangle$ of the impact parameter in the laboratory frame.
 We use an exponential decay law  for the $\tau$ leptons from 
 $h$-decay  with an average
$\tau$ decay length of $\langle{c\tau_{\tau}}\rangle=87\mu m$.
 If one assumes that the pion is emitted, in the $\tau$ rest fame,
 transversely to the $\tau$ direction of flight, then we obtain
   $\langle|{\bf  n}_\mp|\rangle= 44\mu m$. This estimate
   indicates also the magnitude of the resolution which must be experimentally achieved
  both for the primary vertex and the tracks of the pions.

\begin{figure}[tb]
\includegraphics[height=5.2cm]{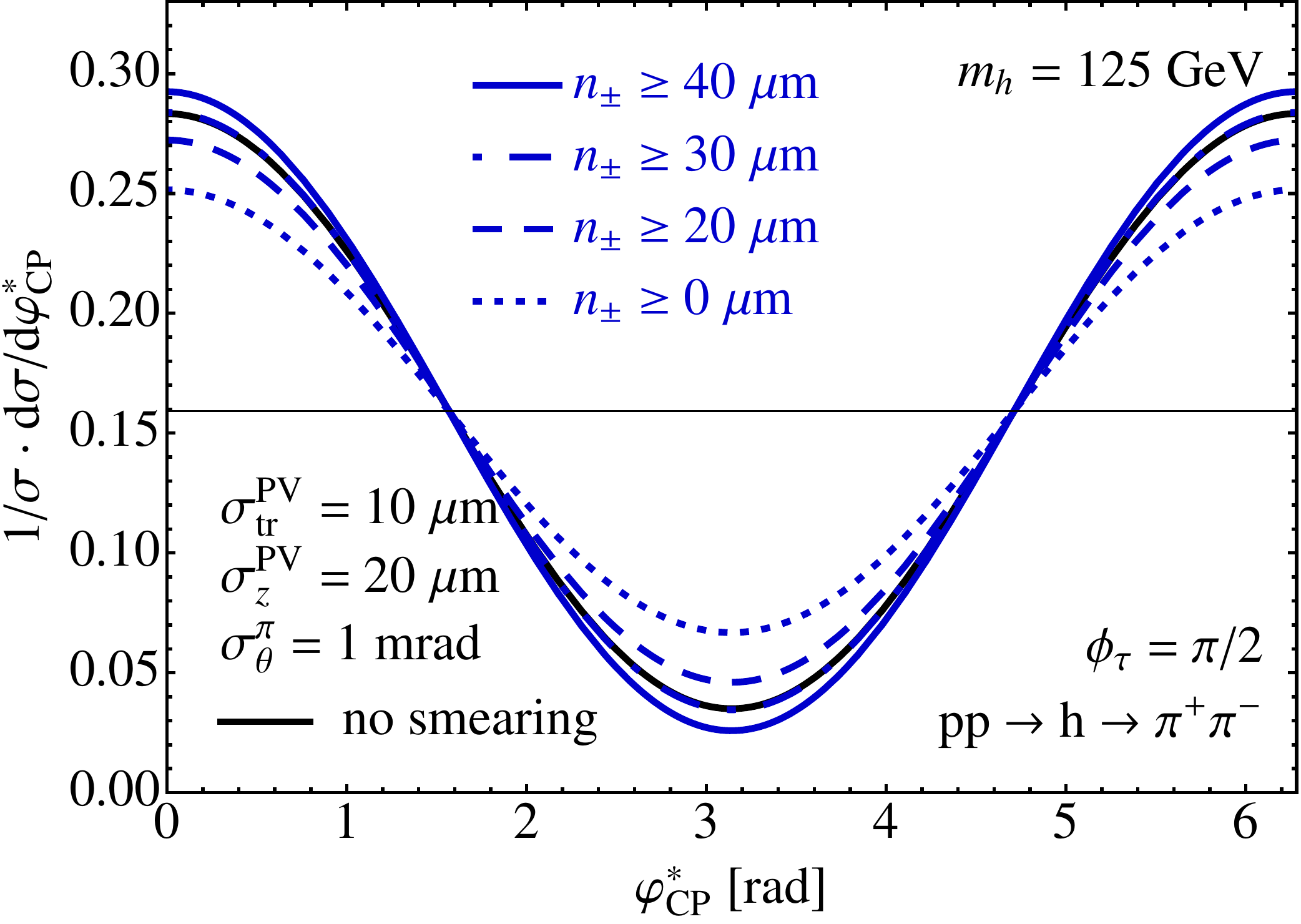}
\hspace{3mm}\includegraphics[height=5.2cm]{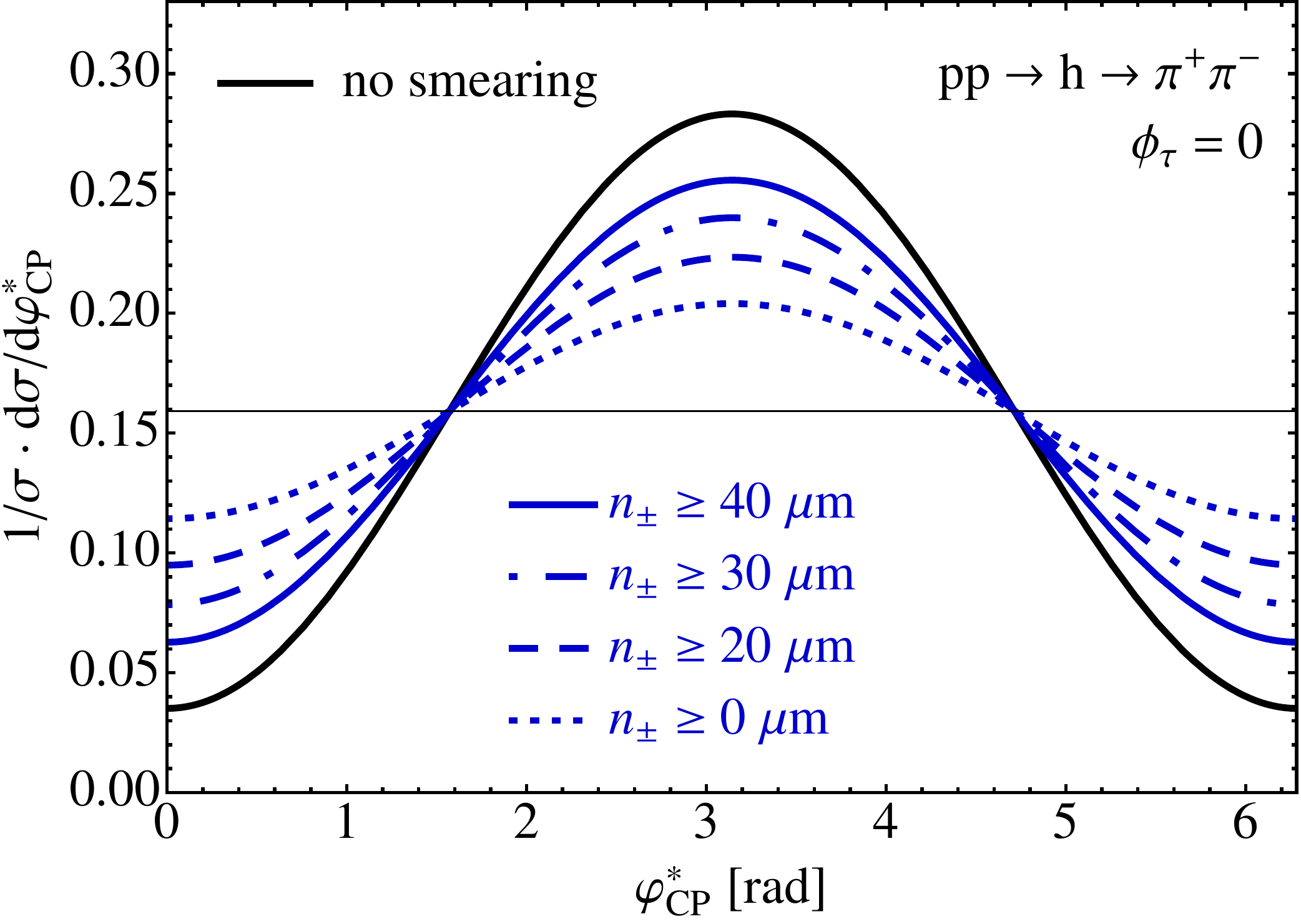}
\caption{LHC ($\sqrt{S}=14$~TeV),  $pp\to
  h\to\tau^{-}\tau^{+}\to\pi^{+}\pi^{-}+2\nu$
 with cuts $p_{T}^{\pi}\ge20$~GeV, $|\eta_{\pi}|\le2.5$.
 The normalized  $\varphi_{CP}^{*}$ distribution, taking measurement
 uncertainties into account, for different minimum cuts on the length
 of the impact parameters $n_\pm$. The left and right plots refer to a CP-odd
 and CP-even Higgs boson, respectively. The horizontal
 lines in both plots are drawn to guide the eye.}
\label{fig:LHC_phi_smearing_A_H} 
\end{figure}

 Taking the smearing of the various quantities into account, with
 standard deviations as specified above, the resulting 
 effects on  the $\varphi_{CP}^{*}$ distribution are shown in 
 Fig.~\ref{fig:LHC_phi_smearing_A_H}, left, for the $\tau\tau\to\pi\pi$
 decay of a CP-odd Higgs boson  and
 Fig.~\ref{fig:LHC_phi_smearing_A_H}, right, for a CP-even Higgs boson.
  The black solid lines show the distributions
without any smearing. The dotted blue lines  include the  effect of 
smearing using the parameters given above. The asymmetry~\eqref{phiCP_asym} is then strongly
reduced from $A_{NLO}^{\pi\pi}=49.5\%$ to $18\%$ ($A_{LO}^{\pi\pi}=18.2\%$) in the case
of a  CP-even Higgs boson and to $37\%$ ($A_{LO}^{\pi\pi}=37.4\%$)
for a CP-odd Higgs boson.  These asymmetries can be enhanced by
taking into account only events with impact parameter lengths $n_\pm$
above a certain minimum value. For the
  cuts  $n_{\pm}\ge20\mu m$, $n_{\pm}\ge30\mu m$, and $n_{\pm}\ge40\mu
 m$, the resulting $\varphi_{CP}^{*}$ distributions are displayed in 
 Fig.~\ref{fig:LHC_phi_smearing_A_H}.  The associated
 asymmetry $A_{NLO}^{\pi\pi}$ is  $25.7\%,\,32.3\%,\,38.6\%$ in the case of a CP-even
 Higgs boson and  $45.2\%,\,49.8\%,\,53.3\%$ for a CP-odd Higgs boson.
 Our Monte Carlo simulations indicate that the value $n_{\rm min}$ of
  the minimum cut should be of the same size as the largest value
  from  the set  $\{\sigma_{z}^{PV},\sigma_{tr}^{PV},\sigma_{tr}^{\pi}\}$.
 Of course,  the number of events is reduced by a cut on $n_\pm$.
It is important to notice  that for a  CP-even or a  CP-odd Higgs boson, the position
of the maximum of the $\varphi_{CP}^{*}$ distribution, whose true value
  is  at $\varphi_{CP}^{*}=0$, respectively
at $\varphi_{CP}^{*}=\pi$,  is neither affected by the  smearing procedure
 nor by a cut on the impact parameters. \\

\begin{figure}[tb]
\includegraphics[height=5.2cm]{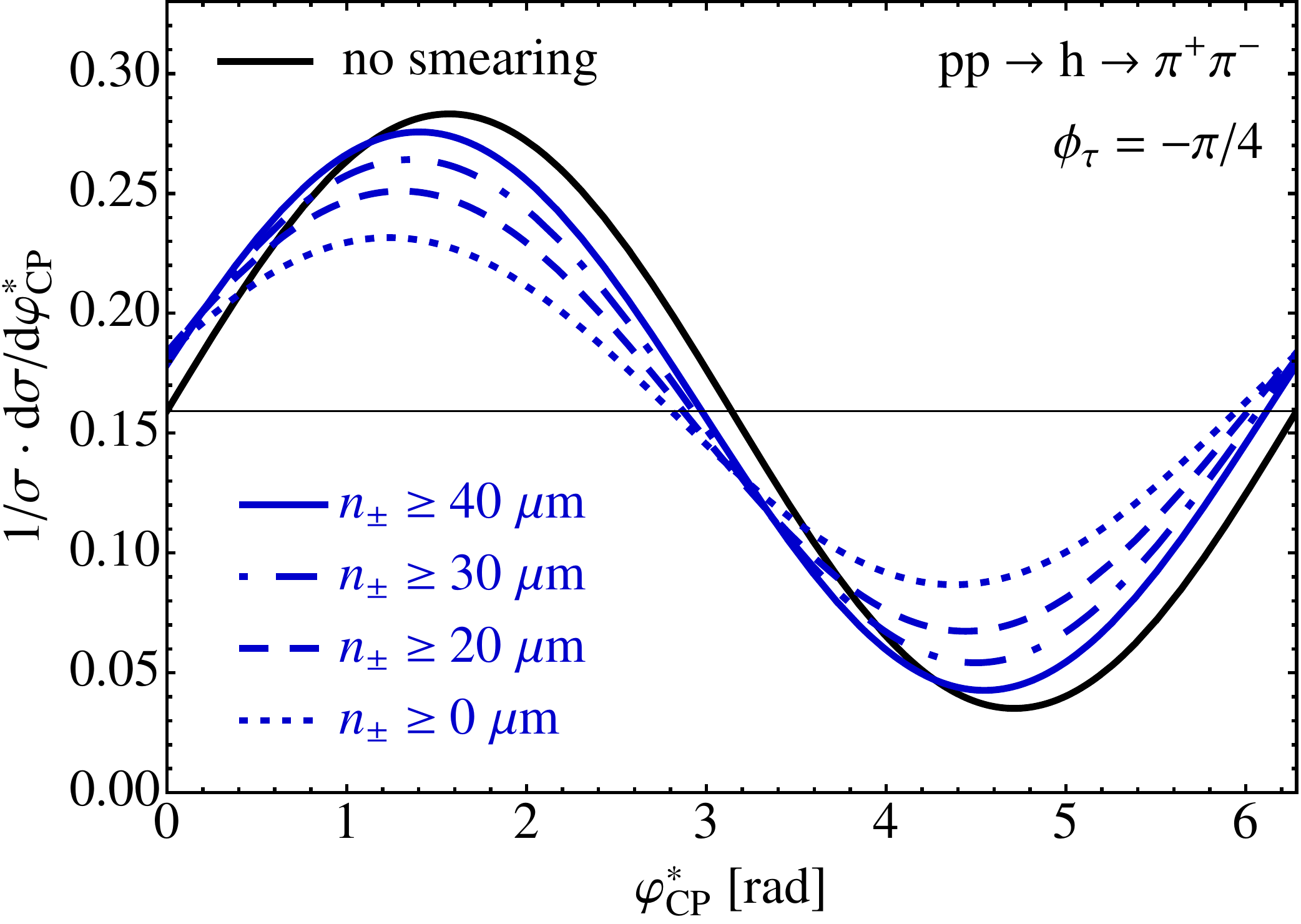}
\hspace{7mm}\includegraphics[height=5.2cm]{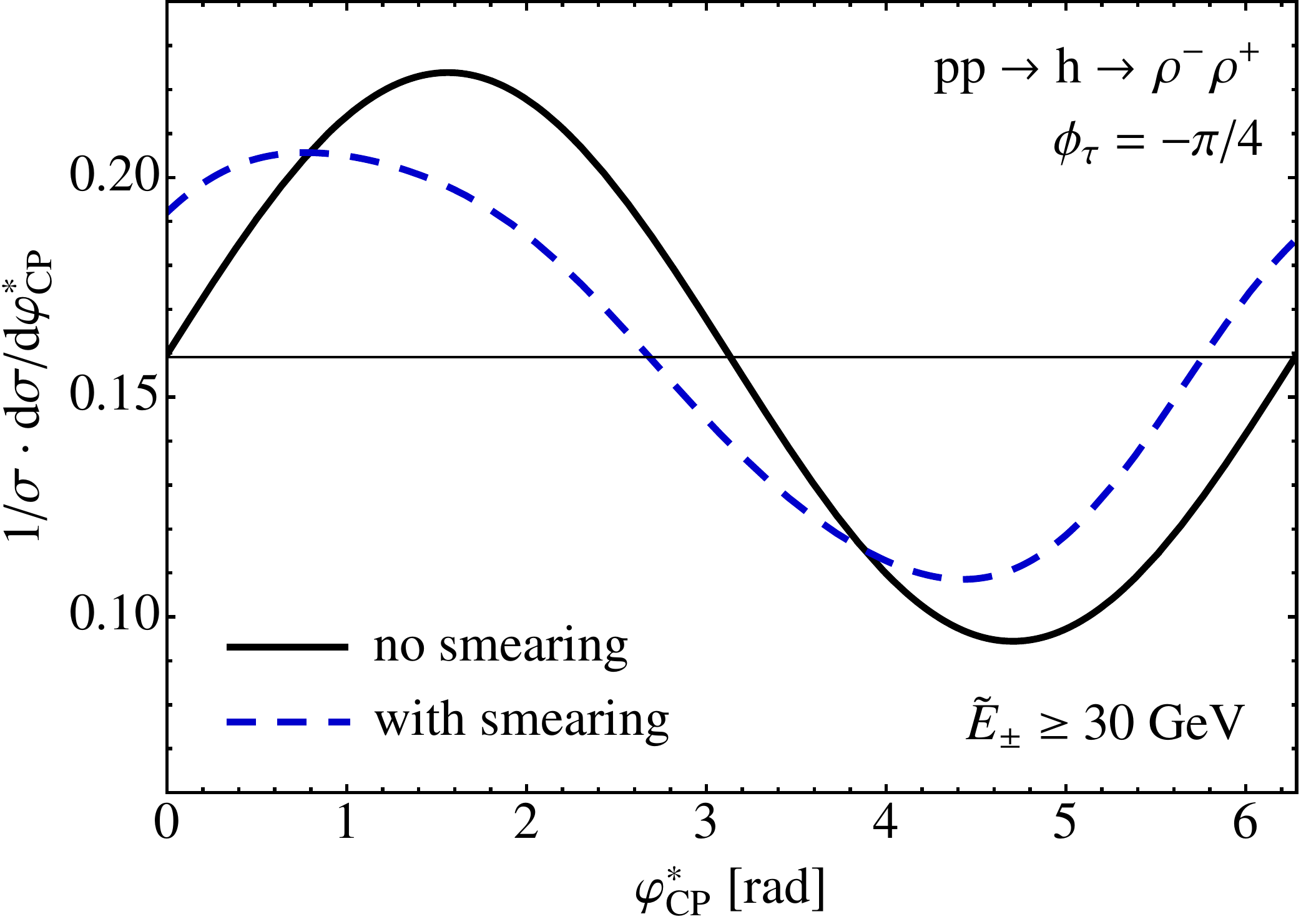}
\caption{LHC ($\sqrt{S}=14$~TeV),  $pp\to
  h +X$, for a CP mixture $h$ with mixing angle
  $\phi_{\tau}=-\pi/4$.
 Left: $h\to\tau^{-}\tau^{+}\to\pi^{+}\pi^{-}$ with
 $p_{T}^{\pi}\ge20$~GeV, $|\eta_{\pi}|\le2.5$. The normalized  $\varphi_{CP}^{*}$ distribution, taking measurement
 uncertainties into account, for different minimum cuts on the length
 of the impact parameters $n_\pm$. 
   Right: $pp\to h\to\tau^{-}\tau^{+}\to\rho^{-}\rho^{+}\to
   \pi^-\pi^+$.
 The normalized  $\varphi_{CP}^{*}$ distribution for events with
 ${\tilde E}_\pm \ge 30$ GeV in the $\tau\tau$ ZMF, without and with
 smearing.  The horizontal
 lines in both plots are drawn to guide the eye. }
 \label{fig:LHC_phi_smearing_CPmix} 
\end{figure}

Next we apply the same smearing procedure and cuts  also to a Higgs boson $h$
being a CP mixture with $\phi_\tau =-{\pi}/{4}$.
 The resulting $\varphi_{CP}^{*}$ distributions are shown 
in the left plot of Fig.~\ref{fig:LHC_phi_smearing_CPmix}.
The asymmetry~\eqref{phiCP_asym}, whose  NLO QCD  value for this decay
mode  is
  $A_{NLO}^{\pi\pi}=49.5\%$ ($A_{LO}^{\pi\pi}=50.2\%$)
is reduced to $29\%$ ($A_{LO}^{\pi\pi}=29.3\%$) by the  measurement uncertainties
 as specified above. By applying  a minimum cut on both impact
 parameters $n_\pm$  the asymmetry can be enhanced to
 $36.7\%,\,42\%,\,46.6\%$ for $n_{\pm}\ge20\mu m\,,30\mu m,\,40\mu m$.
 More importantly, however, the position of the maximum of the $\varphi_{CP}^{*}$
distribution turns out to depend on the smearing parameters and on the 
 cut on $n_\pm$.
 In the case of no smearing the position of the
maximum is at $\varphi_{CP,max}^{*}={\pi}/{2}=1.57.$ With smearing
and no cut on $n_{\pm}$ the location of the maximum moves to $\varphi_{CP,max}^{*}=1.24.$
For $n_{\pm}\ge20\mu m\,,30\mu m,\,40\mu m$ the maximum 
 is at $\varphi_{CP,max}^{*}=1.3,\,1.36,\,1.4$. The 
reason for this shift of the maximum is the smearing of the primary vertex. For
larger values of $\sigma_{z}^{PV}$ or $\sigma_{tr}^{PV}$ the reconstructed
PV moves further away from the two  tracks of $\pi^\pm$. Therefore, the angle
between the two impact parameters becomes smaller. This leads  to an
enhancement of the $\varphi_{CP}^{*}$ distribution near
$\varphi_{CP}^{*}\sim 0$ and $\varphi_{CP}^{*}\sim2\pi$.
In the case of smearing, $p_{T}^\pi$ cuts affect  also the  position of the
maximum of the $\varphi_{CP}^{*}$ distribution.
  Because the Higgs mixing angle $\phi_\tau$ is determined from the difference
   between the position $\varphi_{CP,{\rm max}}^{*}$ of the maximum of the
   measured distribution and  $\varphi_{CP}^{*}=\pi$ (cf. Fig.~\ref{fig:h_pipi_detcuts})
  it is crucial to understand the measurement uncertainties. 
 As will be shown in the
 next section, the measurement of the $\varphi_{CP}^{*}$
distribution for  Drell-Yan $\tau$-pair production  can be used to get a
handle on these uncertainties.

This shift of the maximum of the $\varphi_{CP}^{*}$
distribution, which occurs for a CP mixture,  can become even larger for decay modes
such as $\rho\rho$ or $\rho a_{1}$ if additional cuts on ${\tilde{E}}_{\pm}$
are applied. As an example, we consider the
$h\to\tau^-\tau^+\to\rho^{-}\rho^{+}\to\pi^-\pi^+$ decay channel
 for a CP mixture $h$  with $\phi_{\tau}=-\pi/4$. We apply the
 cuts  ${\tilde{E}}_{\pm}\ge 30$~GeV in order to obtain a large
 asymmetry~\eqref{phiCP_asym}, cf. Sec.~\ref{suse:othertaum}. 
 The solid black curve in 
 Fig.~\ref{fig:LHC_phi_smearing_CPmix}, right, shows
the  associated normalized $\varphi_{CP}^{*}$ distribution without
smearing. Taking measurement uncertainties into account  one obtains
the distribution given by the blue dashed line. Its maximum is 
shifted from $\varphi_{CP,max}^{*}=\pi/2$ to
$\varphi_{CP,max}^{*}\simeq 1$. Notice that  the smeared distribution is raised at $\varphi_{CP}^{*}\sim0$
and $\varphi_{CP}^{*}\sim2\pi$ and lowered at
$\varphi_{CP}^{*}\sim\pi$ as compared to the unsmeared distribution.
 This is due to the smearing of the PV. The PV uncertainty causes the
 same effect on the  $\varphi_{CP}^{*}$ distributions of the
 other $\rho\rho\to\pi\pi$ event categories, e.g. for events 
 with ${\tilde{E}}_{-}\ge 30$~GeV
and ${\tilde{E}}_{+}\le 30$~GeV, and of other $\tau\tau\to a  a'$ decay
modes. At this point we notice that an increase of the uncertainties
of the other parameters discussed above makes the  $\varphi_{CP}^{*}$
distributions flatter.

\subsection{The $\varphi_{CP}^{*}$ distribution for  Drell-Yan
  production of $\tau$ pairs \label{susec:phi_star_CP-Drell-Yan}}

\begin{figure}[tb]
\includegraphics[height=5.2cm]{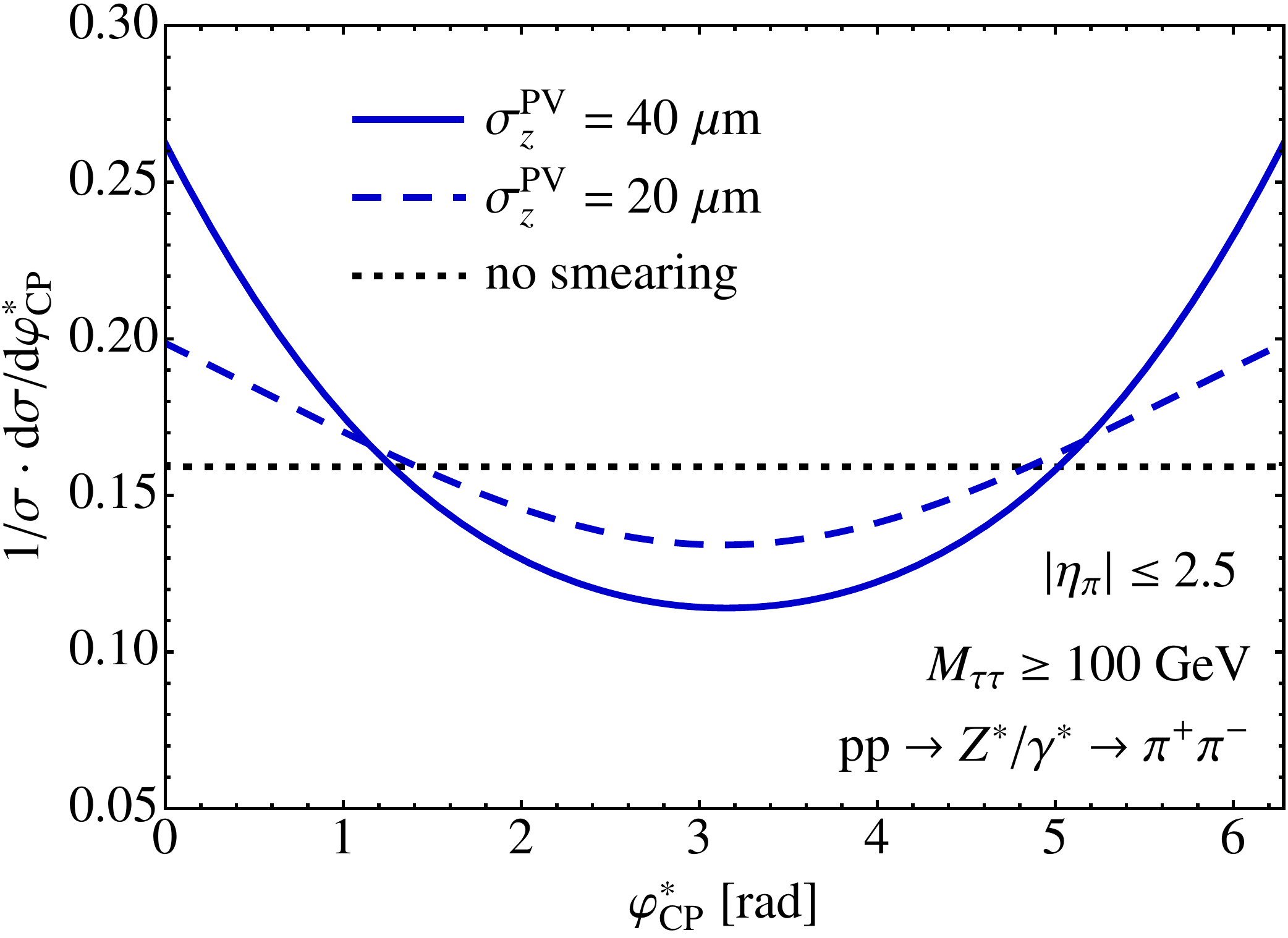}
\hspace{7mm}\includegraphics[height=5.2cm]{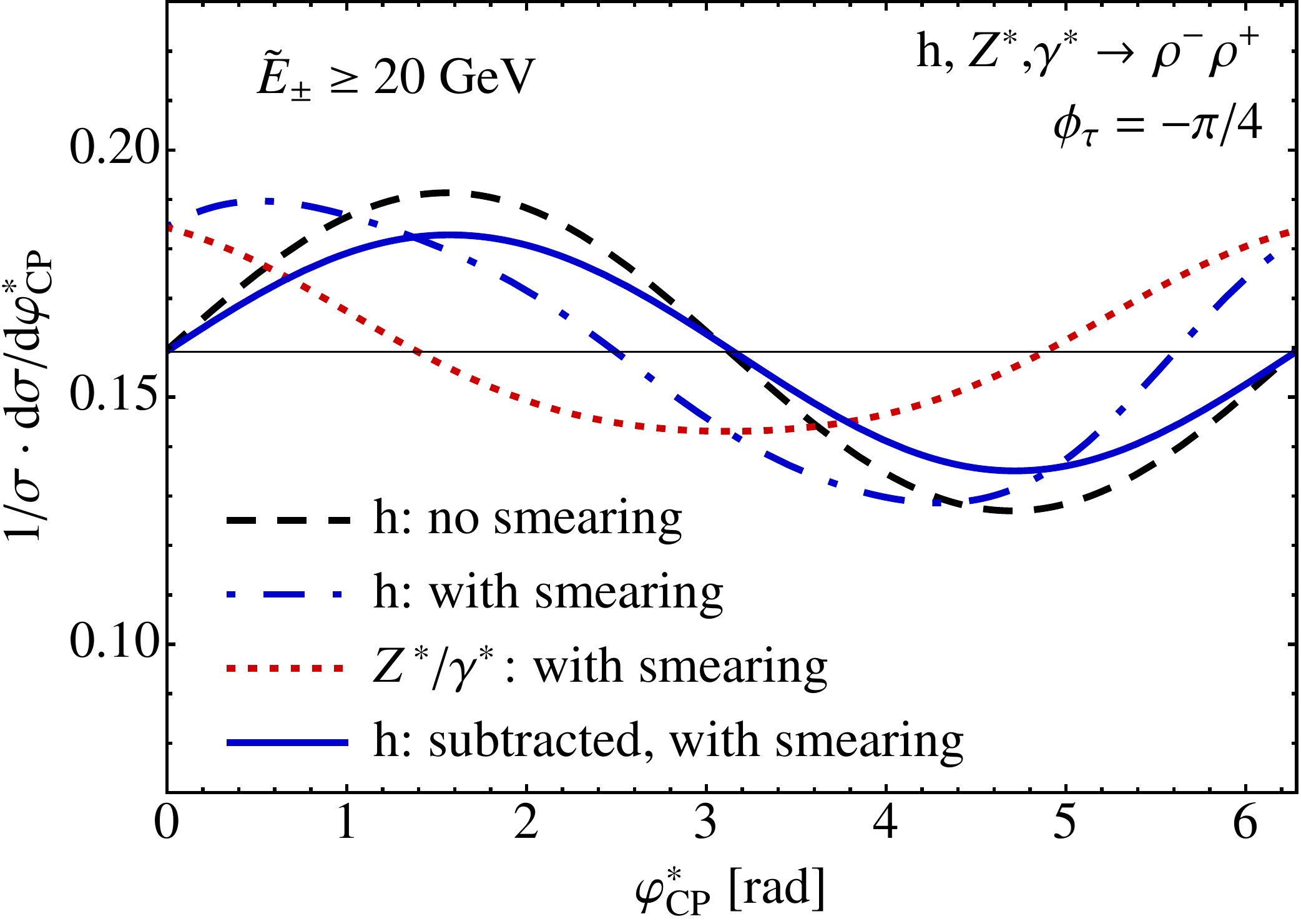}
\caption{Left: $pp\to Z^{*}/\gamma^{*}\to\tau^{-}\tau^{+}\to\pi^{+}\pi^{-}$
 with $p_{T}^{\pi}\ge20$~GeV and $n_{\pm}\ge20\,\mu m$. Normalized
 smeared $\varphi_{CP}^{*}$ distribution for two different values of
 $\sigma_{z}^{PV}$.
 The dotted black line is the prediction without measurement uncertainties.
Right: decays $h, Z^*/\gamma^* \to
\tau^-\tau^+\to\rho^-\rho^+\to\pi^-\pi^+$,  taking $h$ to be a CP mixture
 with $\phi_{\tau}=-\pi/4$. Normalized (un)smeared $\varphi_{CP}^{*}$
  distributions for events with  $\tilde{E}_{\pm}\ge20$~GeV. }
\label{fig:LHC_phi_yZ2pipi_smearing} 
\end{figure}

In this section we investigate the impact of measurement uncertainties on the $\varphi_{CP}^{*}$
distribution  for the Drell-Yan production of $\tau\tau$.
If not stated otherwise we use, as above, for the smeared
distributions the parameters  $\sigma_{z}^{PV}=20\mu m$,
$\sigma_{tr}^{PV}=10\mu m$, $\sigma_{tr}^{\pi}=10\mu m$, $\sigma_{\theta}^{\pi}=1$~mrad
and $\Delta E^{\pi}/E^{\pi}=5\%$. Furthermore, we apply the cut  $n_{\pm}\ge20\,\mu m$
 in the computation of the  distributions of this section. 
For definiteness, we consider the   normalized $\varphi_{CP}^{*}$
distribution  for the 
$pp\to Z^{*}/\gamma^{*}\to\tau^{-}\tau^{+}\to\pi^{+}\pi^{-}+2\nu$
production mode. Without smearing the distribution 
 is given by the dotted black flat line in
 Fig.~\ref{fig:LHC_phi_yZ2pipi_smearing}, left,
 while the effect of the PV uncertainty, simulated with 
 $\sigma_{z}^{PV}=20\mu m$ and with  $\sigma_{z}^{PV}=40\mu m$, results
  in the distribution shown by the dashed and solid blue line, respectively.
  The shape of these curves can be understood as follows. 
If the measurement uncertainty of the PV becomes larger, the distance of
the reconstructed PV to each of the tracks of the charged pions
$\pi^\pm$ 
increases. This results in a smaller angle between the two reconstructed impact parameter vectors.
This, in turn, enhances the region of $\varphi_{CP}^{*}\sim0$ and
$\varphi_{CP}^{*}\sim2\pi$ in the  $\varphi_{CP}^{*}$ distribution.
 On the other hand we found that  larger values of $\sigma_{tr}^{\pi}$,
$\sigma_{\theta}^{\pi}$ and $\Delta E^{\pi}$  decrease the curvature
of the smeared distribution.

An important result of our simulation of the smeared  normalized $\varphi_{CP}^{*}$
distributions for $h\to\tau\tau$ and $Z^{*}/\gamma^{*}\to\tau\tau$ 
 is that they are both enhanced (for $h$ of arbitrary CP nature) at
 $\varphi_{CP}^{*}\sim0$ and $\varphi_{CP}^{*}\sim2\pi$ as compared to
 the respective unsmeared distribution -- an effect which is due to 
    the finite experimental resolution  of the primary vertex. Based on
    this result we suggest the following procedure to obtain a 
     $\varphi_{CP}^{*}$ distribution for the signal reactions 
   $h\to\tau\tau\to a a'$, with which these distortions can be
 eliminated to a large extent.  We assume that a clean data sample
 of Drell-Yan $\tau$ pair events can be recorded at the LHC.  
 One measures  the normalized $\varphi_{CP}^{*}$ distribution for this
 sample and subtracts it from the distribution  measured with the
 $\tau\tau$ events in the signal region, $M_{\tau\tau} \sim 125$ GeV.

We exemplify this proposal for $h, Z^*/\gamma^* \to
\tau^-\tau^+\to\rho^-\rho^+\to\pi^-\pi^+$, taking $h$ to be a CP mixture with
$\phi_{\tau}=-\pi/4$. Fig.~\ref{fig:LHC_phi_yZ2pipi_smearing},
right, shows the unsmeared and smeared $\varphi_{CP}^{*}$
distributions for $h$ decay and the smeared distribution for
$Z^*/\gamma^*$ decay. Subtracting the latter distribution from the
  smeared distribution for $h$ decay and adding the flat line
  $\sigma^{-1}d\sigma/d\varphi_{CP}^{*}=1/(2\pi)$, one obtains
 the solid blue curve. The maximum of this corrected signal
  distribution is at  $\varphi_{CP}^{*,max}=\pi/2$ where it should be.
 This illustrates that with this procedure, one gets rid of the
 distortions caused by the PV measurement uncertainties  to a large
 extent. 

\begin{figure}[tb]
\includegraphics[height=5.2cm]{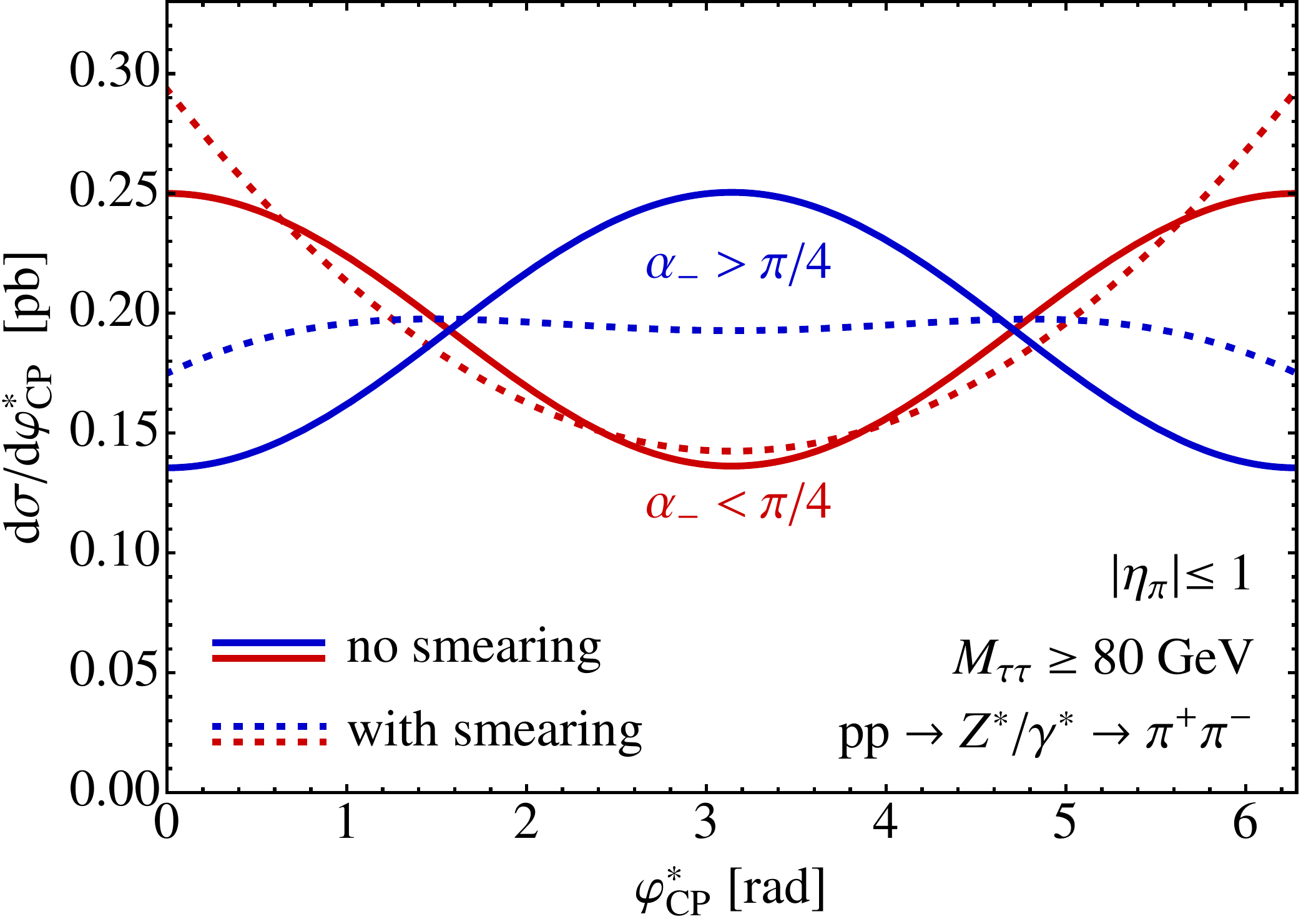}
\caption{$pp\to Z^{*}/\gamma^{*}\to\tau^{-}\tau^{+}\to\pi^{+}\pi^{-}+2\nu$.
 (Un)smeared  $\varphi_{CP}^{*}$
distributions for events with $\pi^-$ being nearly coplanar  ($\alpha_{-}<\pi/4$)  
 and events with $\pi^-$ being nearly perpendicular  ($\alpha_{-}>\pi/4$).}
\label{fig:LHC_phi_yZ2pipi_smearing_alpha} 
\end{figure}

Finally, we compute for $pp\to
Z^{*}/\gamma^{*}\to\tau^{-}\tau^{+}\to\pi^{+}\pi^{-}+2\nu$
 the smeared  $\varphi_{CP}^{*}$
distribution for events with $\pi^-$ being nearly coplanar  ($\alpha_{-}<\pi/4$)  
 and events with $\pi^-$ being nearly perpendicular  ($\alpha_{-}>\pi/4$),
 cf. Sec.~\ref{sec:dsigmaZ_LHC}. The  distribution for
 $\alpha_{-}>\pi/4$  gets significantly distorted by the smearing, as  shown in
 Fig.~\ref{fig:LHC_phi_yZ2pipi_smearing_alpha}. 
 As already discussed in  Sec.~\ref{sec:dsigmaZ_LHC} we propose to
 measure these two distributions as a means to calibrate the signal distribution.

\subsection{Estimate of $\Delta\phi_{\tau}$ \label{susec:Mixing_Angle_precision}}

In this section we estimate the statistical uncertainty
$\Delta\phi_{\tau}$
 with which the mixing angle $\phi_{\tau}$ can be determined from the smeared  $\varphi_{CP}^{*}$
distributions of the $h\to\tau \tau$ decay modes, taking into account
the  $\tau$ decay channels (\ref{taulept}) - (\ref{taua1LT}). As
discussed above, the asymmetry $A^{aa'}$ in Eq.~\eqref{phiCP_asym} is
a measure of the statistical uncertainty
$\Delta\phi_{\tau}$ for each decay channel $aa'$, for a given number
of events\footnote{The analysis of the various decays $h\to\tau\tau \to$
  charged-prongs made here is analogous to our investigation in \cite{Berge:2013jra}, where
 Higgs-boson production in $e^+e^-$ collisions by $e^+e^-\to Z h$ was
 considered. In the present analysis, the cuts on the energies of the pions and charged leptons are made in the Higgs-boson rest frame. }.

  For each $\tau$-decay
 mode the asymmetry \eqref{phiCP_asym}
 is calculated by generating the smeared $\varphi_{CP}^{*}$
distribution of the Higgs-boson signal and of the $Z^{*}/\gamma^{*}$ background
with our Monte Carlo program, using the smearing parameters $\sigma_{z}^{PV}=20\mu m$,
$\sigma_{tr}^{PV}=10\mu m$, $\sigma_{tr}^{\pi}=10\mu m$, $\sigma_{\theta}^{\pi}=1$~mrad,
$\Delta E^{\pi}/E^{\pi}=5\%$, the cut $n_{\pm}\ge20\,\mu m$ on the
length of the impact parameters
and the cut $M_{\tau\tau}\ge100$~GeV. 
 Furthermore, for the leptonic
decay modes we apply  the charged-lepton cuts $p_{T,l}\ge20$~GeV and
$|\eta_{l}|\le2.5$. For the hadronic decay modes the  cuts 
$p_{T,\tau}\ge20$~GeV and $|\eta_{\tau}|\le2.5$ are used, which
 approximate roughly  corresponding cuts on the hadronic $\tau$ jets
 used in experiments.
 As discussed in the last section, we correct the 
   normalized Higgs-boson  $\varphi_{CP}^{*}$ distribution,
  for each decay channel,   
 by subtracting the   normalized $\varphi_{CP}^{*,Z\gamma}$
distribution and adding the  flat distribution $1/(2\pi)$. The
resulting distribution is then  reweighted in order that it
  is properly normalized.
   From this distribution  we calculate
the signal asymmetry $A_{S}$. It is given for the final-state event categories
 `hadron-hadron'  (had-had), `lepton-hadron' (lep-had), and
 `lepton-lepton' (lep-lep)
 in column 2 of Table~\ref{tab:fin_Asymmetry_estimate}. The asymmetry
for signal plus background is then obtained by $A_{S+B}=A_{S}\times {S}/{(S+B)}$.

\begin{table}
\begin{tabular}{|c|c|c|c|c|c|}
\hline 
$\tau\tau$ decay channel & $\quad A_{S}\,\,[\%]\quad$ &
$\quad\frac{S}{S+B}\quad$ & $\quad A_{S+B}\,\,[\%]\quad$ &
$\quad{\sigma_{n_{\pm}\ge20\,\mu m}}/{\sigma}\quad$ & ${\rm events}/fb$\tabularnewline
\hline 
\hline 
had-had & $13.2$ & $0.5$ & $6.6$ & $0.58$ & $1.16$\tabularnewline
\hline 
lep-had & $9.0$ & $0.5$ & $4.5$ & $0.63$ & $1.26$\tabularnewline
\hline 
lep-lep & $7.0$ & ${1}/{3}$ & $2.3$ & $0.61$ & 1.22\tabularnewline
\hline 
~~combined~~ &  &  & $4.85$ & $0.61$ & $3.66$\tabularnewline
\hline 
\end{tabular}\caption{Asymmetries and signal reduction for the hadron-hadron, lepton-hadron, and lepton-lepton
decay modes. The estimate of \textquotedbl{}events/fb\textquotedbl{} includes background events.
\label{tab:fin_Asymmetry_estimate} }
\end{table}
 
In order to estimate the number of events including the background that may be available for
  $\varphi_{CP}^{*}$ measurements at the LHC (14 TeV),  we assume
 for the hadron-hadron and lepton-hadron decay channels the ratio 
ratio $S/B=1$ and\footnote{We have extracted these numbers from the
  results of the  ATLAS experiment for the hadron-hadron and lepton-hadron
  channels in $h\to\tau\tau$ at the LHC (8 TeV) \cite{ATLtauconf}.
 The actual number of events
 at the LHC (14 TeV) will be  higher -- however, the $S/B$
ratio will be worse.} 
  $S+B=2\: {\rm events}/fb$  \cite{ATLtauconf}. For the 
lepton-lepton decay modes we assume  $S/B=1/2$ and $S+B=2\:
{\rm events}/fb$. The resulting ratios $S/(S+B)$ and asymmetries
   $A_{S+B}$ are given in 
  column 3 and 4 of Table~\ref{tab:fin_Asymmetry_estimate}. 
 Next we calculate the factor  $R_{n}={\sigma_{n_{\pm}\ge20\,\mu
     m}}/{\sigma}$ by which the respective signal cross section 
is reduced by a cut on the impact parameters.
 These factors are given  in column 5 of Table~\ref{tab:fin_Asymmetry_estimate}.
 The number of events/fb is then given by $(S+B)\times R_{n}$,
 cf. column 6 of Table~\ref{tab:fin_Asymmetry_estimate}. 

\begin{figure}[tb]
\includegraphics[height=5.8cm]{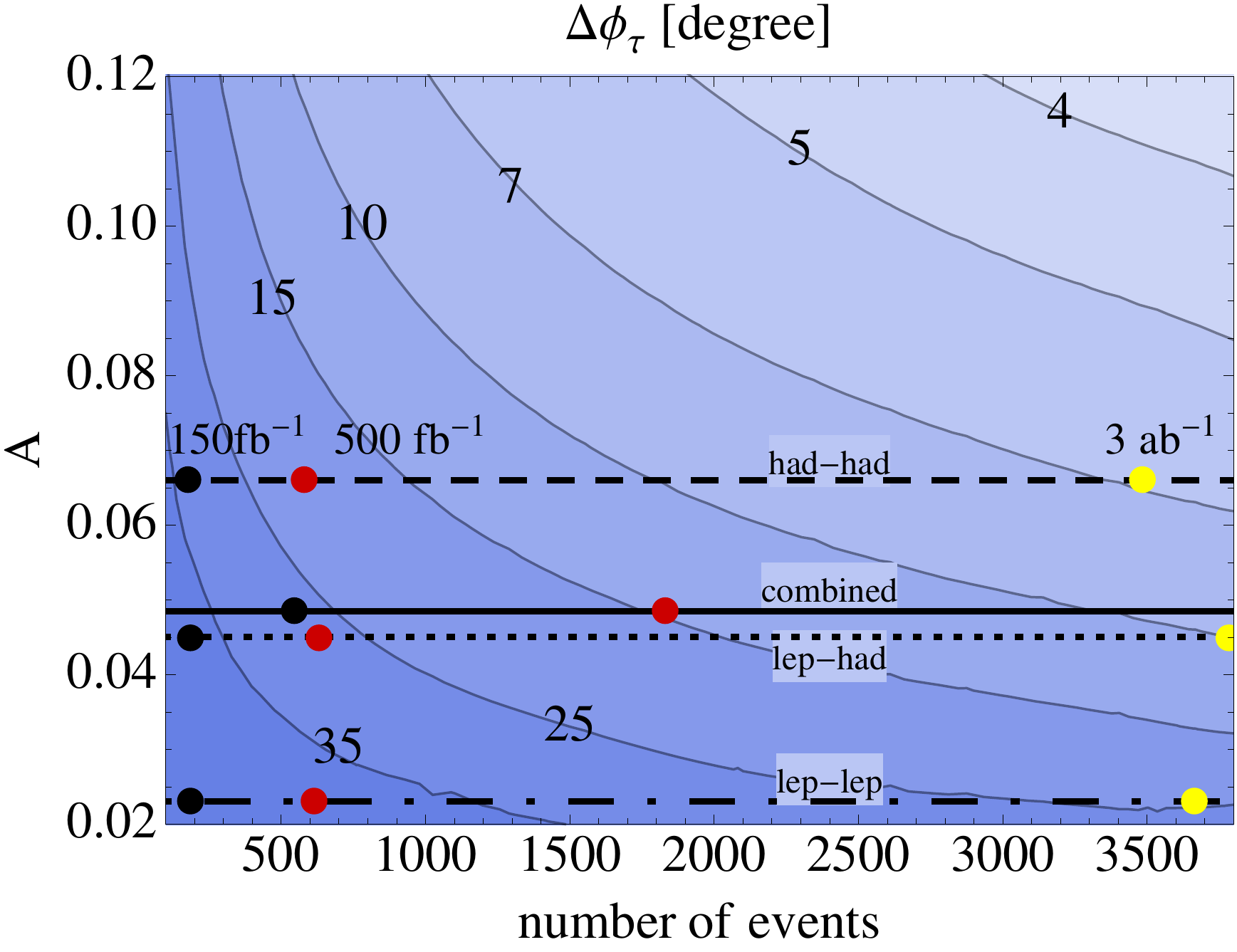}
\caption{LHC ($\sqrt{S}=14$~TeV). Estimated
   statistical  uncertainty $\Delta\phi_{\tau}$ of the Higgs mixing
angle $\phi_\tau$ as a function of the number of events (signal +
background). The horizontal lines display the asymmetries $A_{S+B}$  of
the three event categories given in
 Table~\ref{tab:fin_Asymmetry_estimate}. The yellow dot on the solid black line is
not shown, it corresponds to 11000 events.}
\label{fig:LHC_etsimated_precision} 
\end{figure}

With these values of the  asymmetry  $A_{S+B}$ and number of events/fb
 for the different event categories introduced above, we  
 estimate the statistical uncertainty
$\Delta\phi_{\tau}$ in the following way~\cite{Desch:2003rw}.
 We choose some value of the Higgs mixing angle, for example
 $\phi_{\tau}=-\pi/4$, 
 and  generate the corresponding
differential $\varphi_{CP}^{*}$ distribution using 20 bins between
$0$ and $2\pi$. We then fit this distribution
with the  function $u \cos(\varphi_{CP}^{*}-2\phi_{\tau})+v$. This is
repeated a sufficiently large number of times ($\sim 1000$ times).
  In this way we obtain a distribution 
 of the values of $\phi_{\tau}$ extracted from these fits. 
 This $\phi_{\tau}$ distribution is fitted with a Gaussian, and we
 take its width as our  estimate of the statistical
 uncertainty $\Delta\phi_{\tau}$.
 The result of this procedure is shown in
 Fig.~\ref{fig:LHC_etsimated_precision}.
 The grey contour lines display  $\Delta\phi_{\tau}$ as a function of
  $A_{S+B}$ and the number of events. 
 The horizontal lines are  the asymmetries  $A_{S+B}$ for the three
 event categories and their combination. Assuming that at  the LHC
 (14 TeV) an integrated luminosity of  $150\: fb^{-1}$,
$500\: fb^{-1}$, and  $3\: ab^{-1}$ will be achieved -- the two latter
numbers are goals for the high-luminosity LHC upgrade
\cite{highlumiLHC} -- the resulting event numbers are sketched in as
black, red, and yellow dots, respectively. The yellow dot on the line
 for the combined asymmetry, which corresponds to 11000 events, is not
 shown.       For these luminosities    our estimate of the
 statistical uncertainty
        $\Delta\phi_{\tau}$ which can be achieved by using
 the  combination of the three event categories  is $27^{\circ}$, $14.3^{\circ}$, and
 $5.1^{\circ}$, respectively.

\section{Summary \label{conclusions}}

We have investigated how precisely the CP nature of the 125 GeV Higgs
boson resonance $h$ can be determined at the LHC (14 TeV) in its decay
to $\tau$ leptons. As to the subsequent $\tau$ decays, we have taken
into account all the major decay modes (\ref{taulept}) -
(\ref{taua1LT}).
Our method for determining the Higgs mixing angle $\phi_\tau$, which
parameterizes the ratio of the reduced pseudoscalar and scalar
Higgs-$\tau$ Yukawa couplings, is based on the distribution of the
angle $\varphi_{CP}^{*}$ defined in  \eqref{phistar_CP}.
 This distribution can be measured for all charged-prong $\tau$ decays
 without having to reconstruct the $\tau^\mp$ rest frames. For
 definiteness, we have considered inclusive Higgs-boson production by
 gluon fusion. The irreducible background from Drell-Yan production of
 $\tau$ pairs was analyzed in detail, in particular its contribution
 to the  $\varphi_{CP}^{*}$ distribution. We have studied by   
 Monte Carlo simulation how measurement uncertainties affect the
 signal and background contributions to this distribution. Based on
 this study we devised a procedure for obtaining a corrected
 distribution of this angle. This procedure eliminates to a large
 extent  the distortions due to the measurement uncertainty of  the Higgs production vertex.
 Moreover, we made a proposal how to use
  the  $\varphi_{CP}^{*}$ distribution of Drell-Yan $\tau$-pair events
 for calibrating the experimental uncertainties. 
 Taking the background from $Z^*/\gamma^*\to\tau\tau$ and
  measurement uncertainties by Monte Carlo simulation
 into account, we found that at the 
  LHC (14 TeV), respectively at the LHC-upgrade, with an integrated 
  luminosity of $150\: fb^{-1}$,
$500\: fb^{-1}$, and  $3\: ab^{-1}$,  the Higgs mixing angle  $\phi_\tau$
     can be determined with a precision  $\Delta\phi_{\tau}$  of $27^{\circ}$, $14.3^{\circ}$, and
 $5.1^{\circ}$.


\section*{Acknowledgments}
 We wish to thank the members of the $M_{\tau\tau}$ working group of
 the  Helmholtz Alliance ``Physics at the Terascale'' for discussions.
The work of W.B. is supported by B.M.B.F., contract 05H12PAE,
 and S.K. is supported by Deutsche Forschungsgemeinschaft through
 Graduiertenkolleg GRK 1675.


\end{document}